\documentclass[]{aastex631}

\usepackage{CJK}

\begin{document}
\begin{CJK*}{UTF8}{gbsn}

\title{PEPSI Investigation, Retrieval, and Atlas of Numerous Giant Atmospheres (PIRANGA). I. \\
The Ubiquity of Fe I Emission and Inversions in Ultra Hot Jupiter Atmospheres}

\author{Sydney Petz}
\altaffiliation{These authors contributed equally to the work.}
\affiliation{Department of Astronomy, The Ohio State University, 4055 McPherson Laboratory, 140 West 18$^{\mathrm{th}}$ Ave., Columbus, OH 43210, USA}

\author[0000-0002-5099-8185]{Marshall C. Johnson} 
\altaffiliation{These authors contributed equally to the work.}
\affiliation{Department of Astronomy, The Ohio State University, 4055 McPherson Laboratory, 140 West 18$^{\mathrm{th}}$ Ave., Columbus, OH 43210, USA}

\author[0000-0002-8823-8237]{Anusha Pai Asnodkar} 
\affiliation{Department of Astronomy, The Ohio State University, 4055 McPherson Laboratory, 140 West 18$^{\mathrm{th}}$ Ave., Columbus, OH 43210, USA}

\author[0000-0002-4531-6899]{Alison Duck} 
\affiliation{Department of Astronomy, The Ohio State University, 4055 McPherson Laboratory, 140 West 18$^{\mathrm{th}}$ Ave., Columbus, OH 43210, USA}

\author[0000-0002-4361-8885]{Ji Wang (王吉)}
\affiliation{Department of Astronomy, The Ohio State University, 4055 McPherson Laboratory, 140 West 18$^{\mathrm{th}}$ Ave., Columbus, OH 43210, USA}

\author{Ilya Ilyin}
\affiliation{Leibniz-Institute for Astrophysics Potsdam (AIP), An der Sternwarte 16, D-14482 Potsdam, Germany}

\author{Klaus G. Strassmeier}
\affiliation{Leibniz-Institute for Astrophysics Potsdam (AIP), An der Sternwarte 16, D-14482 Potsdam, Germany}
\affiliation{Institute of Physics \& Astronomy, University of Potsdam, Karl-Liebknecht-Str. 24/25, D-14476 Potsdam, Germany}



\begin{abstract}

We present high-resolution optical emission spectroscopy observations of the ultra hot Jupiters (UHJs) TOI-1431~b and TOI-1518~b using the PEPSI spectrograph on the LBT. 
We detect emission lines from Fe~\textsc{i} with a  significance of 5.68 $\sigma$ and 7.68 $\sigma$ for TOI 1431~b and TOI-1518~b, respectively.
We also tentatively detect Cr~\textsc{i} emission from TOI-1431~b at $4.32\sigma$. For TOI-1518~b, we tentatively detect Ni~\textsc{i}, Fe~\textsc{ii}, and Mg~\textsc{i} at significance levels ranging from $3-4\sigma$. 
Detection of emission lines indicates that both planets possess temperature inversions in their atmospheres, providing further evidence of the ubiquity of stratospheres among UHJs.
By analyzing the population of hot Jupiters, we compare models that predict the distribution of planets in the temperature-gravity space, and find a recent global circulation model suite from \cite{Roth2024} provides a reasonable match to the observed onset of inversions at $T_{\mathrm{eq}}\sim2000$ K.
The ubiquity of strong Fe~\textsc{i} emission lines among UHJs, together with the paucity of detections of TiO, suggest that atomic iron is the dominant optical opacity source in their atmospheres and can be responsible for the inversions.

\end{abstract}



\section{Introduction}

Ultra hot Jupiters (UHJs), short-period giant planets with dayside temperatures in excess of $\sim$2200 K, are some of the most observationally accessible exoplanets. Their hot, bright daysides allow direct detection of light emitted from the planet with high-resolution spectroscopy \citep[e.g.,][]{Pino2020,Yan2022,Brogi23,PETS-II,
PETS-IV}. Many of these planets also reside around hot, bright, A- and early F-type stars, resulting in multiple targets around $V<10$ stars for which the highest signal-to-noise ratio data can be obtained. Over the past few years, high-resolution emission spectroscopy has been published for nine bright UHJ targets. Those with optical spectra uniformly find strong Fe~\textsc{i} emission lines indicating the presence of a temperature inversion in the atmosphere \citep[e.g.,][]{Pino2020,Yan2020,Nugroho2020,Borsa2022,PETS-III,Hoeijmakers2024}, while CO emission lines are seen in the infrared \citep[e.g.,][]{Yan2022,Yan2023}. Other species are seen in both the optical and IR, including atomic metals like Fe~\textsc{ii} and Cr~\textsc{i} and molecules like H$_2$O and OH \citep[e.g.,][]{PETS-III,Brogi23}. In this paper, we add two more to the number of UHJs characterized with high-resolution emission spectroscopy: TOI-1431~b and TOI-1518~b, bringing the total sample size to eleven.

\subsection{TOI-1431 b}
\label{TOI-1431b}

TOI-1431~b (also known as MASCARA-5~b), is an exoplanet discovered by the \textit{Transiting Exoplanet Survey Satellite} \citep[\emph{TESS};][]{Ricker2015} along with the Multi-site All-Sky CAmeRA \citep[MASCARA;][]{Talens17} and confirmed to be a planet by \cite{Addison2021}. Orbiting a bright ($V=8.03$) Am star with an effective temperature of 7690 K at a period of 2.65 days, TOI-1431b has an equilibrium temperature of 2370 K. 
Due to the high temperatures of this UHJ, as well as its short orbital period, TOI-1431~b is a prime candidate for emission spectroscopy to investigate its atmospheric chemistry.

\cite{Stangret2021} used high resolution transmission spectroscopy to search for a variety of atomic and molecular species in the atmosphere of TOI-1431~b, but were unable to find any evidence of atmospheric features. They ascribed the lack of detections to this planet's relatively high surface gravity ($\log g = 3.54$ and consequent small atmospheric scale height. \cite{Stangret2021} also found that the orbit of TOI-1431~b is highly inclined with respect to the stellar rotation.
We investigate the atmospheric chemistry of this planet using high resolution emission spectroscopy for the first time.

\subsection{TOI-1518 b}
\label{TOI-1518b}

TOI-1518~b is another ultra hot Jupiter discovered in \textit{TESS} data by \cite{Cabot2021}. It orbits a bright ($V=8.95$), rapidly rotating F0 host star every 1.9 days, resulting in an equilibrium temperature of 2492 K. It is on a highly inclined, retrograde orbit, and nodal precession of the orbit was recently detected through photometric follow-up observations \citep{Watanabe24}.
\cite{Cabot2021} also detected absorption from both Fe~\textsc{i} and Fe~\textsc{ii} during transit. TOI-1518~b was identified by \cite{Hord24} as one of the five best UHJs for emission spectroscopy, but to our knowledge no observations of the dayside atmosphere have yet been published.

\subsection{Atmospheric Inversions}

Emission spectroscopy is sensitive to the pressure-temperature (P-T) profile in the photosphere of exoplanets \citep[e.g.,][]{Coulombe2023}. Decreasing temperature with altitude results in absorption lines in the spectrum, whereas an increasing temperature (a \textit{temperature inversion}, also known as a stratosphere) results in emission lines. These can be detected with either high-resolution \citep[e.g.,][]{Birkby13,Brogi23} or low-resolution space-based spectroscopy \citep[e.g.,][]{Mansfield2018,Fu22a}.

An inversion requires the presence of a strong opacity source in the upper atmosphere, which absorbs incoming stellar radiation and heats the atmosphere. The most commonly invoked inversion agents for hot Jupiters are metal oxides like TiO and VO \citep[e.g.,][]{Fortney08}. Observational searches for these species, however, have been controversial and had only limited success using high resolution spectroscopy \citep[e.g.,][]{Nugroho2017,Serindag2021,Prinoth22,PETS-II}. 
However, there have been few claimed TiO detections in low resolution \citep{Cont21}.
Many other species have been proposed as possible inversion agents, including other metal hydrides and oxides such as SiO, FeH, or CaH; atomic species such as Fe~\textsc{i} or Mg~\textsc{i}; or continuum opacity due to the bound-free opacity of H$^-$ \citep{Lothringer18,GandhiMadhusudhan19}. Optical high-resolution spectroscopy is optimal for searching for these species, as they have many lines in this wavelength range. 

\cite{Beatty17-KELT1,Beatty17-Kep13} found that two high-gravity objects lacked inversions (Kepler-13~Ab: $\log g=4.0$, KELT-1~b: $\log g=4.7$), even in the same equilibrium temperature range where lower-gravity objects have inverted atmospheres. They proposed that this is due to the presence of an atmospheric cold-trap related to the surface gravity, with higher-gravity objects possessing more efficient cold traps that would remove TiO and VO from their atmospheres and prevent the formation of a temperature inversion.

In this paper we search for the presence of atmospheric inversions in TOI-1431~b and TOI-1518~b, both of which have moderate surface gravity, using high-resolution spectroscopy (\S\ref{sec:observations}), as well as catalog the atomic and molecular species, including potential inversion agents, detectable in their atmospheres (see \S\ref{sec:methodology} for the methodology and \S\ref{sec:results} for the results). Subsequently (\S\ref{sec:litanalysis}), we investigate the distribution of the presence of inversions (or the lack thereof) across the parameter space of hot Jupiters, and attempt to link this to the inversion agent(s).

\section{Observations}
\label{sec:observations}

\begin{figure}
\centering
\includegraphics[width=0.375\textwidth]{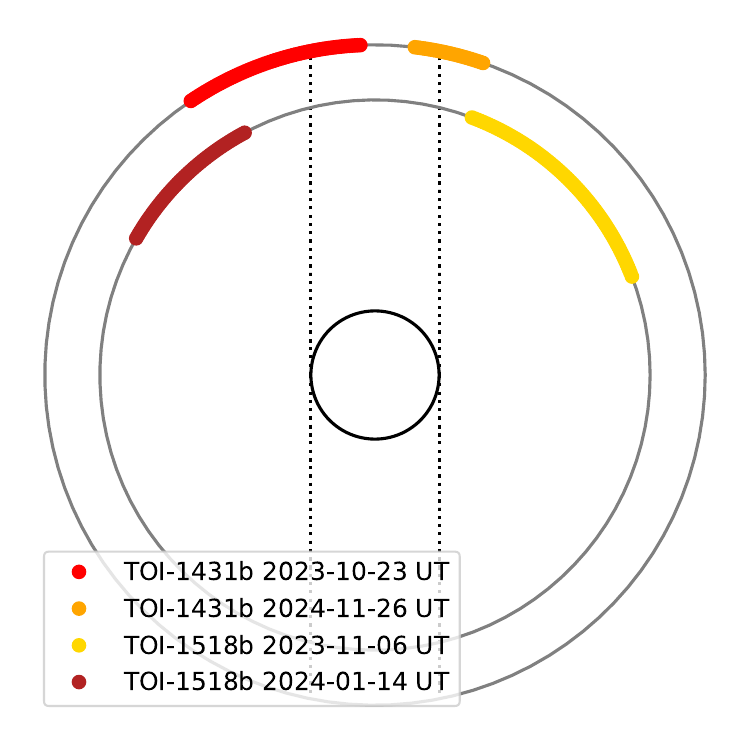}
\includegraphics[width=0.5\textwidth]{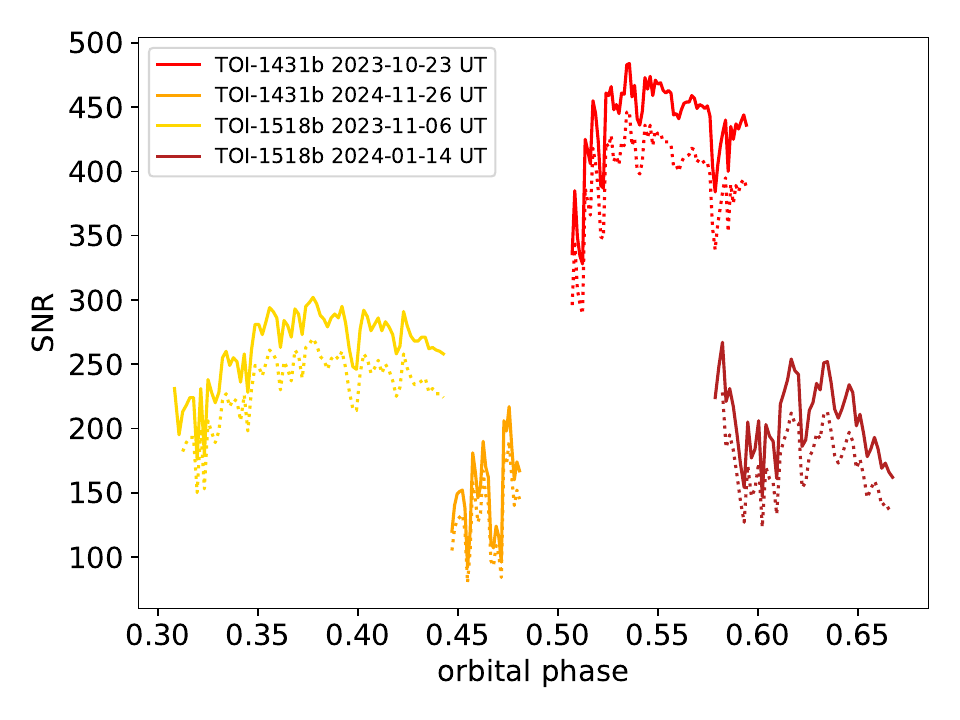}
\caption{Left: Phase coverage of the observations used in this work. The solid inner circle shows the stellar surface, while the middle and outer circles show the orbits of TOI-1518~b and TOI-1431~b, respectively, to scale. The observer is off the bottom of the page, and the vertical dashed lines show the line of sight, such that transit occurs in the lower intersection of the dashed lines and planetary orbits, and the secondary eclipse in the upper intersection. The planets orbit counter-clockwise. The colored points show the portions of the orbit where we obtained data. Right: signal-to-noise ratio of our observations as a function of orbital phase. The PEPSI red and blue arms are shown as the solid and dotted lines, respectively. The SNR values shown are, for each spectrum, the $95^{\mathrm{th}}$ quantile per-pixel signal-to-noise ratios.}
\label{fig:phase-coverage}
\end{figure}

We obtained time-series spectra of TOI-1431 and TOI-1518 for a duration of 2-7 hours per night using the Potsdam \'Echelle Polarimetric and Spectrographic Instrument \citep[PEPSI;][]{Strassmeier15} on the $2\times8.4$ m Large Binocular Telescope (LBT) located on Mt. Graham, Arizona, USA. We observed both targets on two nights, both covering both pre- and post-secondary eclipse; we show the phase coverage in Fig.~\ref{fig:phase-coverage}. We used the PEPSI 200$\mu$m fiber, giving a resolving power of $R=130,000$, and used Cross Dispersers (CD) III and V, providing wavelength coverage of 4800-5441 \AA\ and 6278-7419 \AA\ in the blue and red channels, respectively. We used an exposure time of 250 seconds in each arm for each target. Further details of the observations are given in Table~\ref{tab:obslog}. The PEPSI CCDs were upgraded during the 2023 summer shutdown, and one of the LBT's two secondary mirrors was re-aluminized, resulting in comparable SNR ($\sim$200-400) at a shorter exposure length (250 seconds) with respect to our previous observations of a brighter star \citep[300 seconds for KELT-20, $V=7.58$;][]{PETS-II,PETS-IV}. However, the 2024-01-14 UT TOI-1518 observations were of lower SNR than the 2023-11-06 UT dataset on the same planet due to poorer observing conditions, and the 2024-11-26 data on TOI-1431~b data were of both lower SNR and short duration, and have only a minimal impact upon our results (right panel of Fig.~\ref{fig:phase-coverage}).

The environmental control system that typically keeps the pressure and temperature around the PEPSI optical bench stable was disabled for the TOI-1431 observations due to a software fault. This resulted in some drift in the wavelength solution over the course of the observations; however, we were able to correct for this drift using the Fabry-Perot etalon simultaneous wavelength calibration system. There was no obvious effect on the recovery of the planetary signal from this drift, which amounted to less than 200 m s$^{-1}$, much smaller than the resolution of the instrument and the tens of km s$^{-1}$ motion of the planetary lines over the course of the observations. We therefore do not expect any significant impact upon our results for TOI-1431~b.

The data were reduced using the SDS4PEPSI pipeline \citep{Ilyin2000,Strassmeier18}. The pipeline performs standard reduction steps, and produces an output 1-dimensional, wavelength-calibrated, continuum-normalized spectrum where the different orders and the traces from the two unit telescopes have been combined into a single spectrum. The pipeline also calculates the variance of each pixel, which we use to propagate uncertainties through the remainder of the analysis.

\begin{deluxetable*}{llcccccc}
\tablecaption{Log of Observations\label{tab:obslog}}
\tablewidth{0pt}
\tablehead{
Target & Date (UT) & $N_{\mathrm{spec}}$ & Exp. Time (s) & Airmass Range & Phases Covered & SNR$_{\mathrm{blue}}$ & SNR$_{\mathrm{red}}$
}
\startdata
TOI-1431 b & 2023-10-23 & 68 & 250 & 1.09 - 1.60 & 0.507 - 0.594 & 396 & 438 \\
TOI-1431~b & 2024-11-26 & 27 & 250 & 1.16 - 1.48 & 0.447 - 0.481 & 132 & 152 \\
TOI-1518 b & 2023-11-06 & 74* & 250 & 1.21 - 1.43 & 0.311 - 0.445 & 232 & 264 \\
TOI-1518 b & 2024-01-14 & 50* & 250 & 1.28 - 2.04 & 0.581 - 0.670 & 173 & 208 \\
\enddata
\tablecomments{$N_{\mathrm{spec}}$ is the number of spectra obtained on that night. Exp. time is the exposure time in seconds. SNR$_{\mathrm{blue}}$ and SNR$_{\mathrm{red}}$ are the nightly average of the $95^{\mathrm{th}}$ quantile per-pixel signal-to-noise ratios in the blue and red arms, respectively.
*We excluded the first two blue-arm spectra from each TOI-1518 dataset, as well as the first red-arm spectrum from 2023 Nov. 6, due to solar contamination.} 
\end{deluxetable*}

\section{Methodology}
\label{sec:methodology}

Our methodology is largely the same as was used in \cite{PETS-II} and \cite{PETS-IV}. In short, we first model out the telluric lines in the red-arm spectra using the MOLECFIT package \citep{Smette15,Kausch15}; although MOLECFIT performs well for correcting weak and moderately strong telluric lines, strong lines and bands are very difficult to correct and we exclude regions of the spectra strongly affected by tellurics from further analysis. We used a restricted wavelength range for the MOLECFIT models as compared to our previous work, choosing narrower ranges with fewer stellar lines: 6293-6298 \AA, 6307-6314 \AA, 6474-6483 \AA, 6509-6517 \AA, 7102-7110 \AA, and 7358-7374 \AA\ (all wavelengths given in the observatory rest frame). This was important for the relatively slowly rotating TOI-1431; with a $v\sin i$ of only 6 km s$^{-1}$ \citep{Addison2021}, TOI-1431 is much more narrow-lined than either KELT-20 \citep{PETS-II,PETS-IV} or TOI-1518. There is therefore a greater risk of MOLECFIT confusing stellar for telluric lines. We found that using these same wavelength ranges also resulted in a better fit (as measured by eye) to the telluric spectrum 
for TOI-1518, so we adopted them for both stars. 

After telluric correction, we interpolate all of the time-series spectra from each PEPSI arm from a single night onto a common wavelength scale and create a median-combined spectrum, which is then removed from each of the time series spectra. Next, we run the SYSREM algorithm \citep{Tamuz05} on the two-dimensional residuals in wavelength-time space in order to remove any imperfectly-removed telluric and stellar lines, and any instrumental systematics that might be present. We use the methodology proposed by Spring \& Birkby (submitted) to halt SYSREM once removal of successive systematics fails to improve to residuals by more than 1 part in $10^{-4}$. This avoids the pitfalls of optimizing SYSREM based on the detection significance, which can result in false detections (Spring \& Birkby, submitted).

We then construct a model planetary emission spectrum using \texttt{petitRADTRANS} \citep{Molliere19}. We assume a pressure-temperature (P-T) profile of the form given by \cite{Guillot10}, and include only a single line species in each model in addition to continuum opacity from H$^{-}$, and collisionally-induced opacity from H$_2$-H$_2$ and H$_2$-He interactions. We adopted most planetary and stellar parameters from the discovery papers from the two systems, which are given in Table~\ref{tab:systempars}. We convolved each model spectrum with a model rotationally broadened line profile of the planet, as well as a model instrumental profile.

In order to assure that we correctly shift the data into the stellar rest frame for each target, we measured the stellar radial velocity in the PEPSI frame using the same methodology as in \cite{Pai2022} and \cite{PETS-IV}. Briefly, we use least squares deconvolution as implemented in \cite{Kochukhov2010} and \cite{Wang2017}, with a \texttt{Spectroscopy Made Easy} \citep{ValentiPiskunov1996,ValentiPiskunov2012} template spectrum, to extract the average stellar line profile from each time-series spectrum. We fit a rotationally broadened stellar line profile computed per \cite{Gray} to the extracted line profiles, and use this to extract the stellar RVs. Finally, we fit a Keplerian model for a circular orbit to these RVs, including a zero-point offset which is the best-fit systemic RV in the PEPSI frame. We quote the resulting RV in Table~\ref{tab:systempars}.

Additionally, we utilize new ephemerides for both planets as the original discovery ephemerides are now several years old, 
additional \textit{TESS} data are available, and the use of incorrect ephemerides can provide inaccurate conclusions on atmospheric dynamics \citep[e.g.,]{Monalto2011, Pai2022b, Smith2024}.We fit the \textit{TESS} photometry using \texttt{EXOFASTv2} \citep{EXOFASTv2} in order to derive updated ephemerides. For both planets we were able to detect the secondary eclipses, providing additional leverage of the ephemeris and constraining the eccentricity (which is consistent with zero for both planets, so we assume circular orbits going forward).
The updated ephemerides are listed in Table~\ref{tab:systempars}, and are part of a larger effort which will be presented in Duck et al. (in prep). 

Next, we cross-correlated the model planetary spectrum with the time-series processed residual spectra, shifted each cross-correlation function (CCF) into the planetary rest frame assuming a range of planetary radial velocity (RV) semi-amplitude ($K_P$) values, and combined them to form a final CCF. We estimated the uncertainty in these maps from the standard deviation of regions at $|v_{\mathrm{sys}}|>100$ km s$^{-1}$ from the expected planetary rest frame. We searched for peaks in this CCF map at the expected $K_P$, $v_{\mathrm{sys}}$ of the planet, and consider peaks with $>5\sigma$ to be detections and those with $3-5\sigma$ to be tentative detections which will require more data to confirm.
For TOI-1431, some of our spectra where taken during secondary eclipse. We excluded these spectra from the cross-correlation analysis, but they were used to compute the median spectrum and through the SYSREM analysis.

\begin{figure*}
\centering
\includegraphics{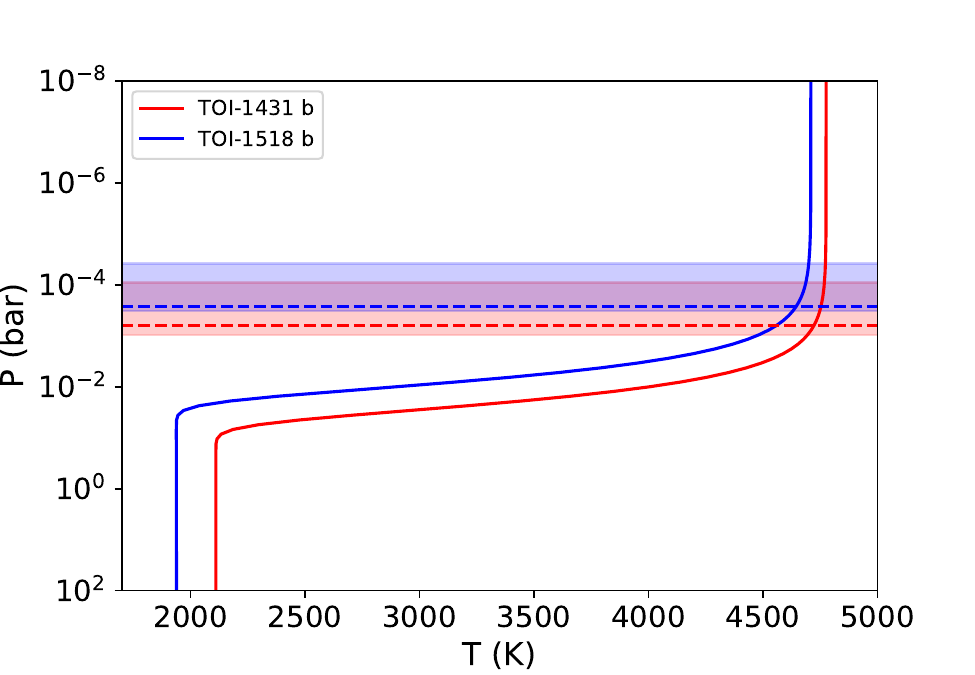}
\caption{Pressure-temperature profiles for TOI-1431~b and TOI-1518~b adopted in this work, shown in red and blue, respectively. The horizontal dashed lines and bands show, in the same per-planet colors, the estimated pressure of peak of the contribution function and 68\% contribution range for the Fe~\textsc{i} emission spectrum, as calculated by \texttt{petitRADTRANS}. We emphasize that these are estimated P-T profiles produced by a manual grid search to maximize SNR and not robust measurements of the profiles.}
\label{fig:PTprofiles}
\end{figure*}

In order to set the P-T profile, we followed the methodology of \cite{PETS-II} by conducting a simple, manual grid search in the parameters $\kappa_{\mathrm{IR}}$ (IR opacity), $\gamma$ (ratio of optical to infrared opacity), $T_{\mathrm{eq}}$, and volume mixing ratio (VMR) of Fe~\textsc{i} and adopted the parameters that resulted in the most significant detection of Fe~\textsc{i} for each system. For each additional species, we vary its VMR in steps of 0.25 dex until we obtain the most significant detection. We emphasize that this is not an exhaustive search, and the resulting P-T profiles and VMR should be taken with caveats; in particular, we have not attempted to account or correct for the distortion of the planetary signal due to the SYSREM process \citep{Gibson2022}, nor to do a rigorous retrieval; the SNR is not a robust likelihood statistic. We will pursue these analyses in a future paper. We show the best-fit P-T profiles in Fig.~\ref{fig:PTprofiles}.

We searched for the following species: Fe~\textsc{i}, Fe~\textsc{ii}, Ni~\textsc{i}, Cr~\textsc{i}, Ti~\textsc{i}, Ti~\textsc{ii}, Mg~\textsc{i}, Si~\textsc{i}, Sc~\textsc{i}, V~\textsc{i}, Mn~\textsc{i}, Sr~\textsc{i}, Ca~\textsc{i}, Ca~\textsc{ii}, TiO, VO, FeH, MgH, CaH, and CrH. We chose these species as they were found to be the most detectable in UHJs similar to TOI-1431~b and TOI-1518~b with PEPSI by \cite{PETS-IV}, plus CrH, which was recently detected in WASP-31~b by \cite{Flagg23}. We used opacities from \url{http://kurucz.harvard.edu/} for the atomic species. For TiO, VO, and FeH, we used opacities from \cite{McKemmish19} ($^{48}\text{Ti}^{16}\text{O}$), \cite{McKemmish16}, and \cite{Wende10}, respectively. For the remaining molecular species, the opacities were obtained from the DACE Opacity Database\footnote{\url{https://dace.unige.ch/opacityDatabase/}} using the Yadin line list \citep{Yadin2012} for MgH, Rivlin line list \citep{Rivlin2015} for CaH, and the MoLLIST line list \citep{Burrows2002} for CrH.

\begin{deluxetable}{ccccc}
\tablecaption{System and Atmospheric Model Parameters\label{tab:systempars}}
\tablewidth{0pt}
\tablehead{
 & \multicolumn{2}{c}{TOI-1431~b} & \multicolumn{2}{c}{TOI-1518~b} \\
Parameter & Value & Source & Value & Source 
}
\startdata
$R_P$ ($R_J$) & $1.49 \pm 0.05$ & \cite{Addison2021} & $1.875 \pm 0.053$ & \cite{Cabot2021} \\
$M_P$ ($M_J$) & $3.12 \pm 0.18$ & \cite{Addison2021} & $<2.3$ ($3\sigma$ limit) & \cite{Cabot2021} \\
$T_{\mathrm{eq}}$ (K) & $2370 \pm 70$ & \cite{Addison2021} & $2492 \pm 38$ & \cite{Cabot2021} \\
$P$ (d) & $2.65023142 \pm 0.00000017$ & Duck et al. (in prep) & $1.9026055 \pm 
0.0000066$ & Duck et al. (in prep) \\
$T_0$ (BJD\_TDB) & $22458712.674837 \pm 0.000072$ & Duck et al. (in prep) & $2459854.41433 \pm 0.00012$ & Duck et al. (in prep) \\
$R_{\star}$ ($R_{\odot}$) & $1.92 \pm 0.07$ & \cite{Addison2021} & $1.950 \pm 0.048$ & \cite{Cabot2021} \\
$M_{\star}$ ($M_{\odot}$) & $1.90_{-0.08}^{+0.10}$ & \cite{Addison2021} & $1.79 \pm 0.26$ & \cite{Cabot2021} \\
$T_{\mathrm{eff}}$ (K) & $7690_{-250}^{+400}$ & \cite{Addison2021} & $7300 \pm 100$ & \cite{Cabot2021} \\
$K_P$ (km s$^{-1}$)* & $187.4 \pm 3.3$ & \cite{Addison2021} & $203.7 \pm 9.9$ & \cite{Cabot2021} \\
$v_{\mathrm{sys}}$ (km s$^{-1}$)$^{\dagger}$ & $-24.902 \pm 0.011$ & This Work & $-11.170 \pm 0.035$ & This Work \\
$v\sin i_P$ (km s$^{-1}$)* & $2.858 \pm 0.096$ & \cite{Addison2021} & $5.01 \pm 0.14$ & \cite{Cabot2021} \\
$\kappa_{\mathrm{IR}}$ & 0.005 & This Work & 0.04 & This Work \\
$\gamma$ & 50 & This Work & 45 & This Work \\
$P_0$ (bar) & 1 & \ldots  & 1 & \ldots \\
$X_{\mathrm{H}_2}$ & 0.7496 & \ldots  & 0.7496 & \ldots \\
$X_{\mathrm{He}}$ & 0.2504 & \ldots  & 0.2504 & \ldots \\
VMR(H$^-$) & $1\times10^{-9}$ & \ldots  & $1\times10^{-9}$ & \ldots \\
$N_{\mathrm{SYSREM,Blue}}$ & 3 & \ldots & 4 & \ldots \\
$N_{\mathrm{SYSREM,Red}}$ & 1 & \ldots & 3 & \ldots \\
\enddata
\tablecomments{* The quoted value is derived from parameters given in the quoted work, but is not explicitly calculated in the reference. $^{\dagger}$ The quoted systemic velocity is measured in the PEPSI frame.}
\end{deluxetable}

\section{Results}
\label{sec:results}

\subsection{The Dayside Spectra of TOI-1431~b and TOI-1518~b}
\label{sec:bothplanets}

We detect a significant Fe~\textsc{i} signals in emission from both TOI-1431~b and TOI-1518~b. The fitting parameters ($v_{\mathrm{sys}}$ and $K_P$) that maximize the detection significance are consistent with those reported in previous literature \citep{Addison2021, Cabot2021}.
The detected emission signals clearly indicate the presence of temperature inversions in the atmospheres of the two planets.
We show the best-fit P-T profiles from the simple grid search, assuming a constant VMR of Fe~\textsc{i} as a function of altitude, in Fig.~\ref{fig:PTprofiles}. Both show a strong inversion as seen in other UHJs \citep[e.g.,][]{Cont21,PETS-II}. 
We show the shifted and combined CCFs for TOI-1431~b in Fig.~\ref{fig:ccfs-toi1431}, and for TOI-1518~b in Figs.~\ref{fig:ccfs-toi1518} and \ref{fig:ccfs-toi1518-2}.
The detected species, significance values, and best-fit VMRs for both planets are listed in Table~\ref{tab:results}, while the P-T profile parameters are listed in Tables~\ref{tab:systempars}. We again emphasize that these parameters have not been produced by a rigorous retrieval, and are likely subject to systematic effects due to simplifying assumptions such as ignoring the distortion of the planetary spectrum due to SYSREM \citep{Gibson2022} or inhomogeneity of the dayside atmosphere. They are unlikely to be correct in detail and should be taken to be representative only; the key point is that both atmospheres must be inverted as the CCFs require the presence of emission lines.

In order to set limits upon the presence of TiO in both planets, we follow the same methodology as used by \cite{PETS-II}. Briefly, we injected a model spectrum using the TiO opacity tables produced by B. Plez and obtained from the petitRADTRANS opacity repository\footnote{\url{https://keeper.mpdl.mpg.de/d/e627411309ba4597a343/}}. We then attempted to retrieve the injected spectrum using a TiO model constructed from the line list of \cite{McKemmish19}, in order to attempt to mimic the effects of line list uncertainty. While such a procedure likely overestimates the detection significance \citep{PETS-II,PETS-IV}, this nonetheless suggests that we should have been able to detect TiO if it was present in both planets with a VMR of $\gtrsim10^{-9}$. This is well below the predicted equilibrium abundances from \texttt{FastChem}, and suggests that TiO is indeed depleted in both planets.

\subsubsection{TOI-1431 b}
\label{sec:1431results}

\begin{figure*}
\centering
\includegraphics[width=0.4\textwidth]{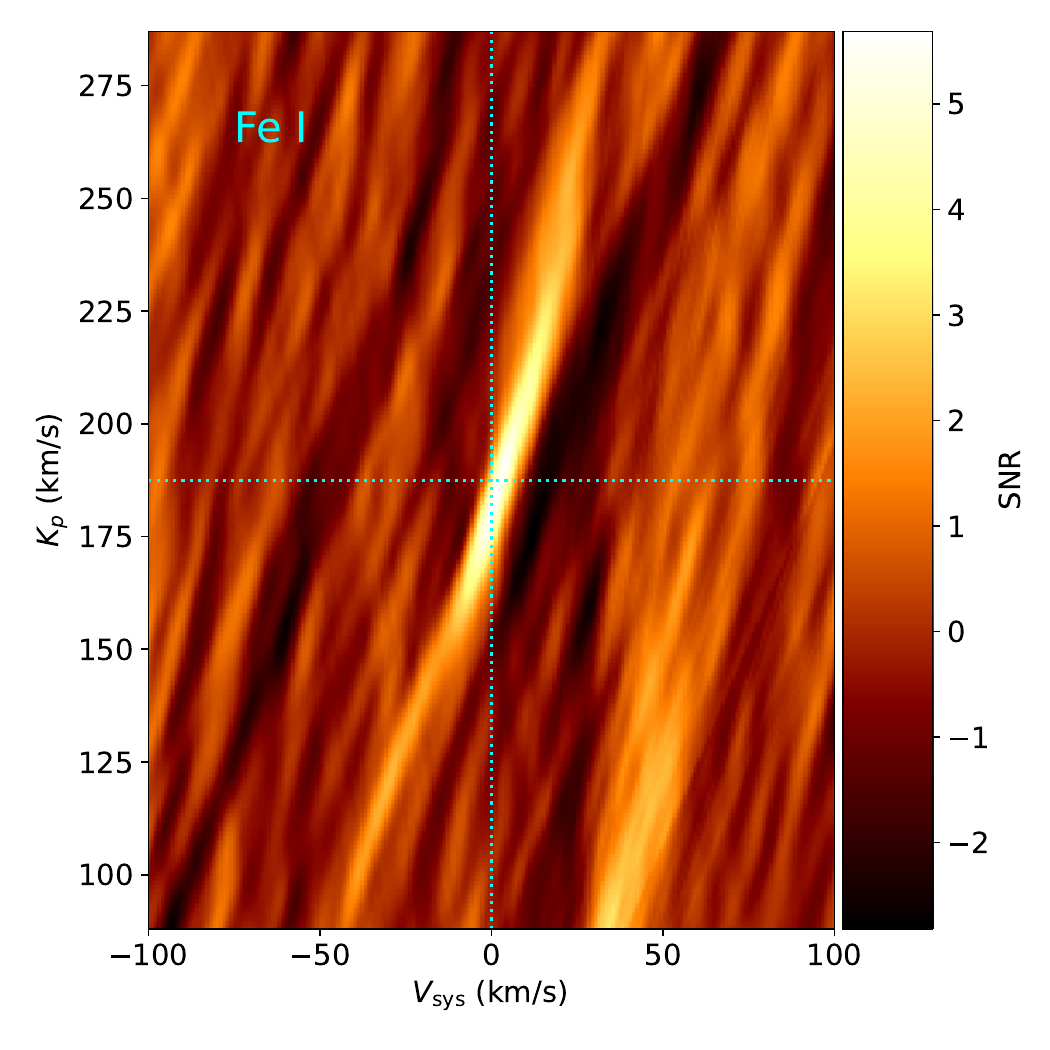}
\includegraphics[width=0.4\textwidth]{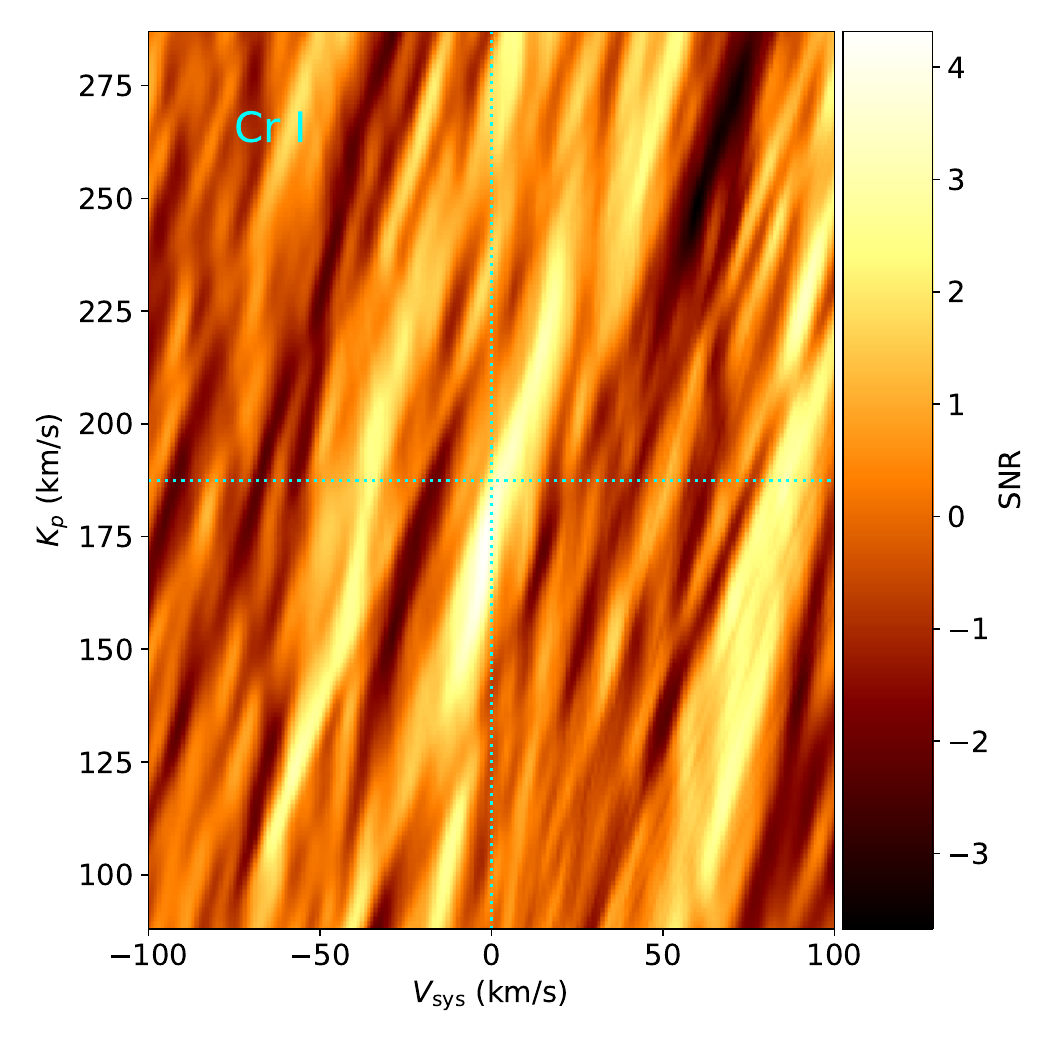}
\caption{CCFs for TOI-1431~b, after being shifted into the planetary rest frame assuming a range of values of the planetary RV semi-amplitude $K_P$ and combined. Fe~\textsc{i} is shown on the left and Cr~\textsc{i} on the right.}
\label{fig:ccfs-toi1431}
\end{figure*}

For TOI-1431~b, only Fe~\textsc{i} exceeds the $5\sigma$ threshold in both the single night of data and after combining both nights of observation. Cr~\textsc{i} is at a tentative detection level of $4.3\sigma$. Combining the two species into a single model, however, gives an $8.04\sigma$ detection for the atmosphere as a whole. 
No other species among those searched rise above a threshold of $3\sigma$.
These are shown in Fig.~\ref{fig:ccfs-toi1431}. More data, including better post-eclipse data, 
will be necessary to search for other species.

\subsubsection{TOI-1518 b}
\label{sec:1518results}

\begin{figure*}
\centering
\includegraphics[width=0.4\textwidth]{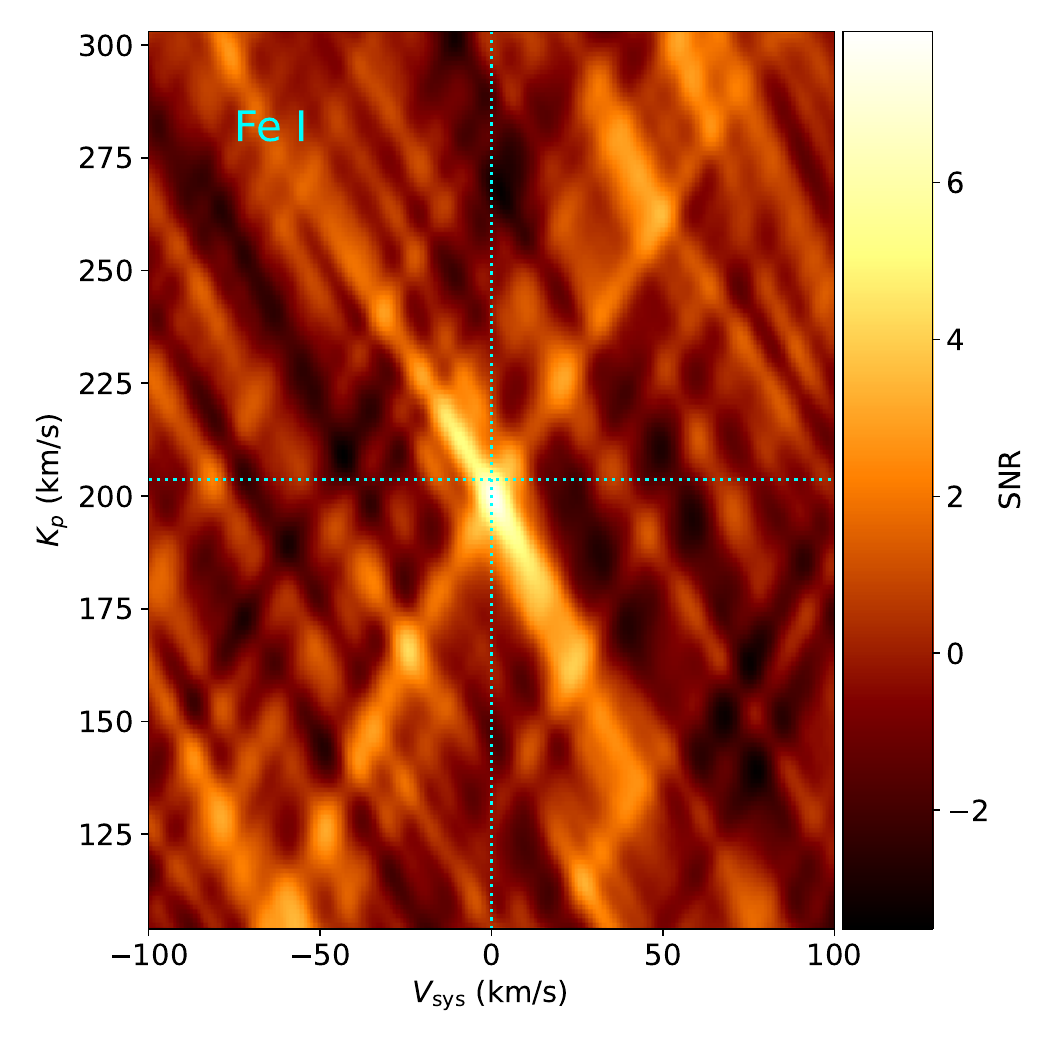} \\
\includegraphics[width=0.4\textwidth]{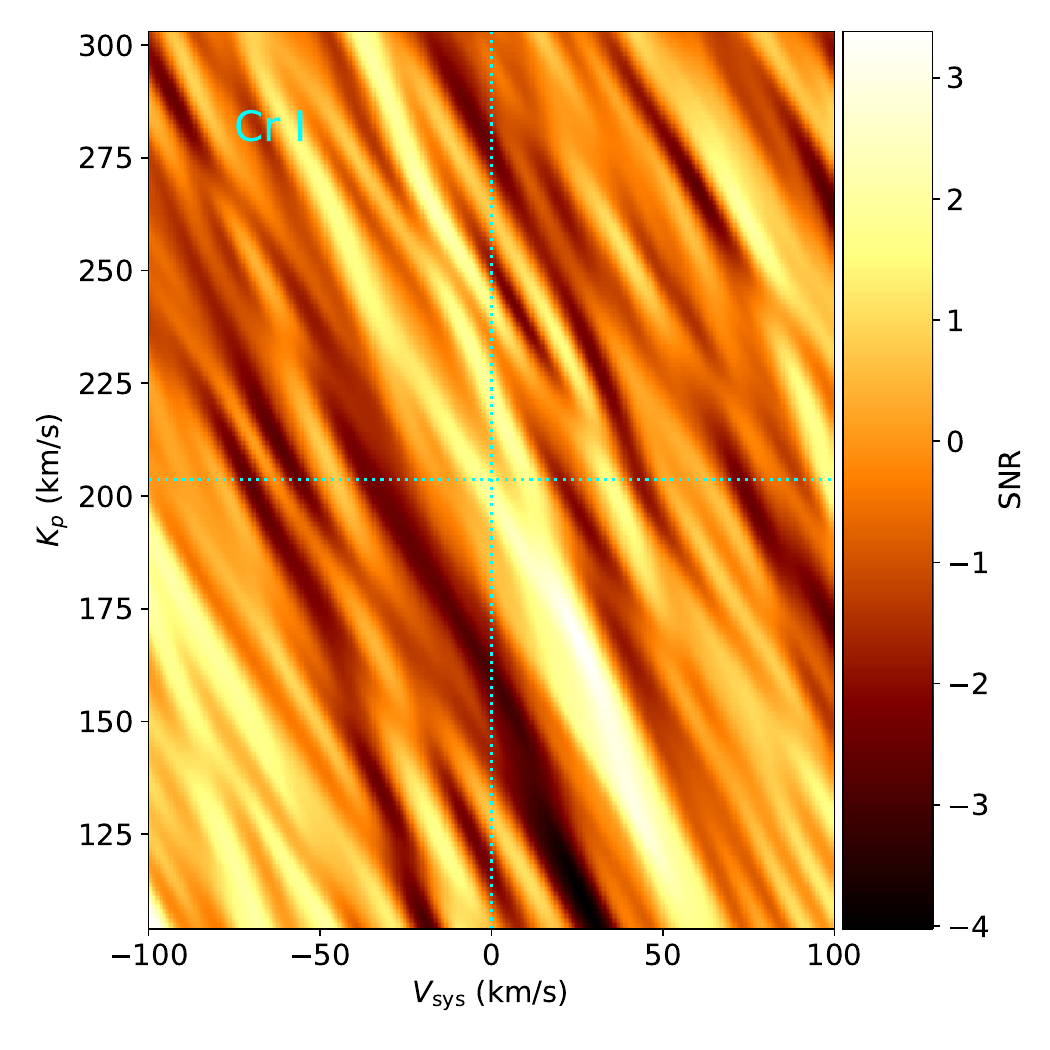}
\includegraphics[width=0.4\textwidth]{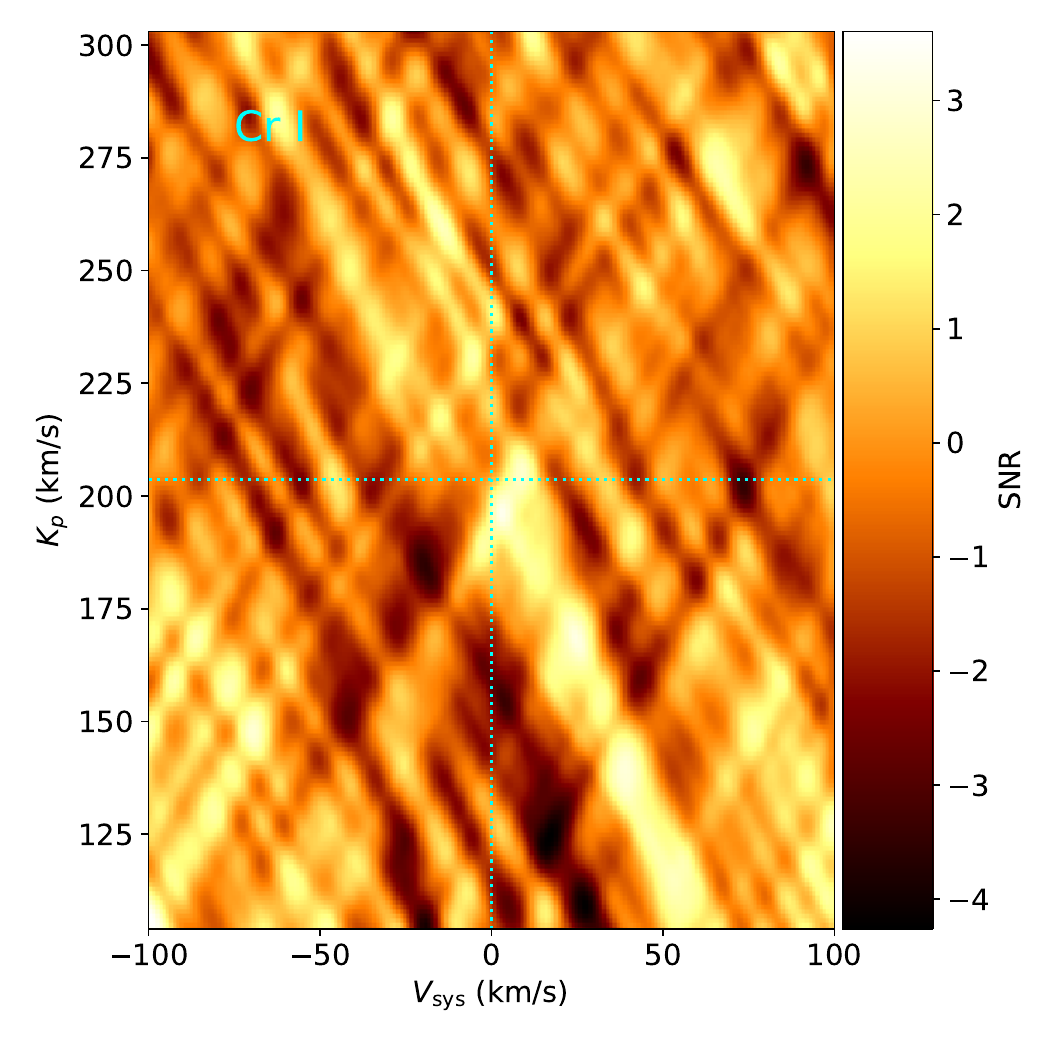} \\

\caption{Shifted and combined CCFs for TOI-1518~b for Fe~\textsc{i} (top), and Cr~\textsc{i} (bottom row). For the bottom row, the left plot shows only the data from 2023 Nov. 6, while the right plot shows both nights. These plots combine the PEPSI red and blue arms. }
\label{fig:ccfs-toi1518}
\end{figure*}

For TOI-1518~b, we make a firm detection ($>5\sigma$) only of Fe~\textsc{i} emission, with a significance of 
7.67$\sigma$ combining both of our nights. We show this detection in the upper left panel of Fig.~\ref{fig:ccfs-toi1518}.

\begin{figure*}
\centering
\includegraphics[width=0.4\textwidth]{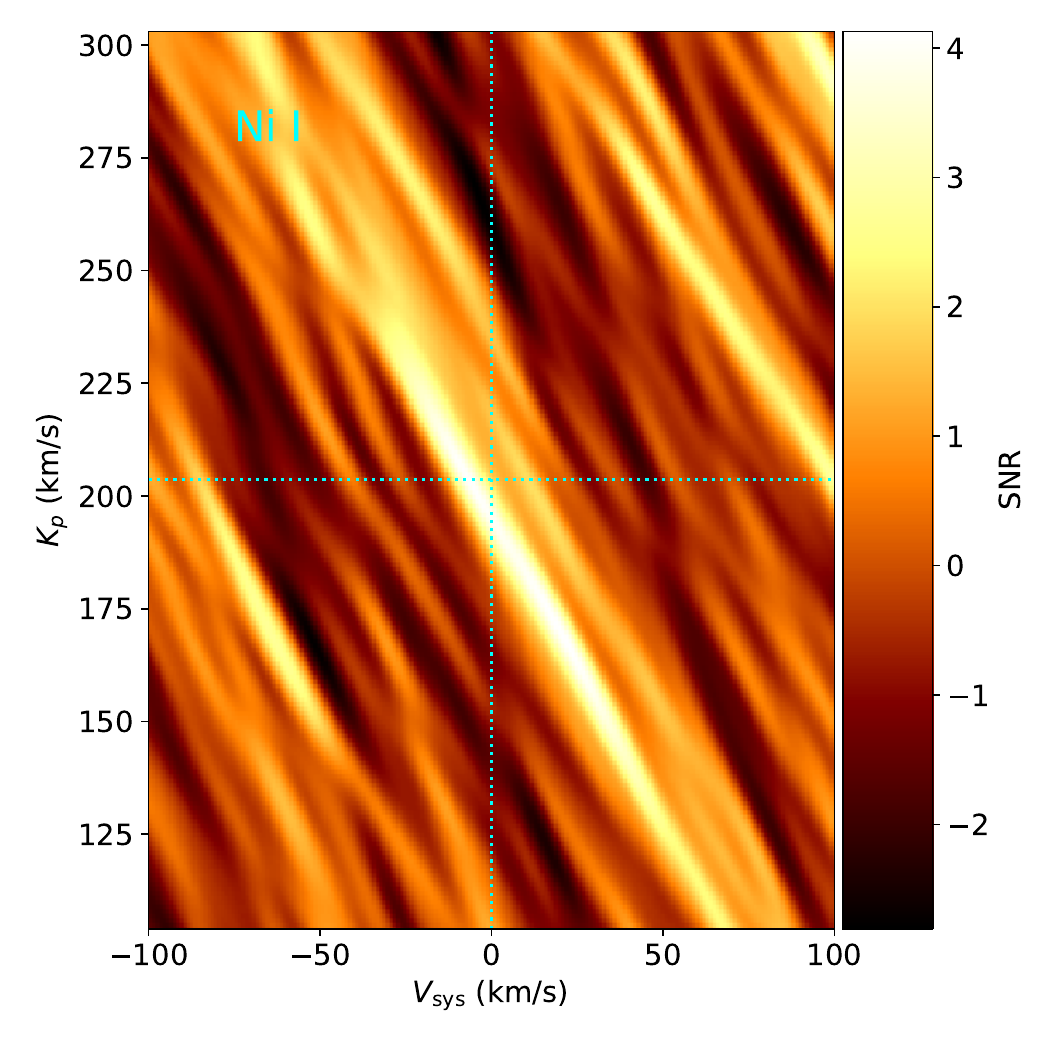}
\includegraphics[width=0.4\textwidth]{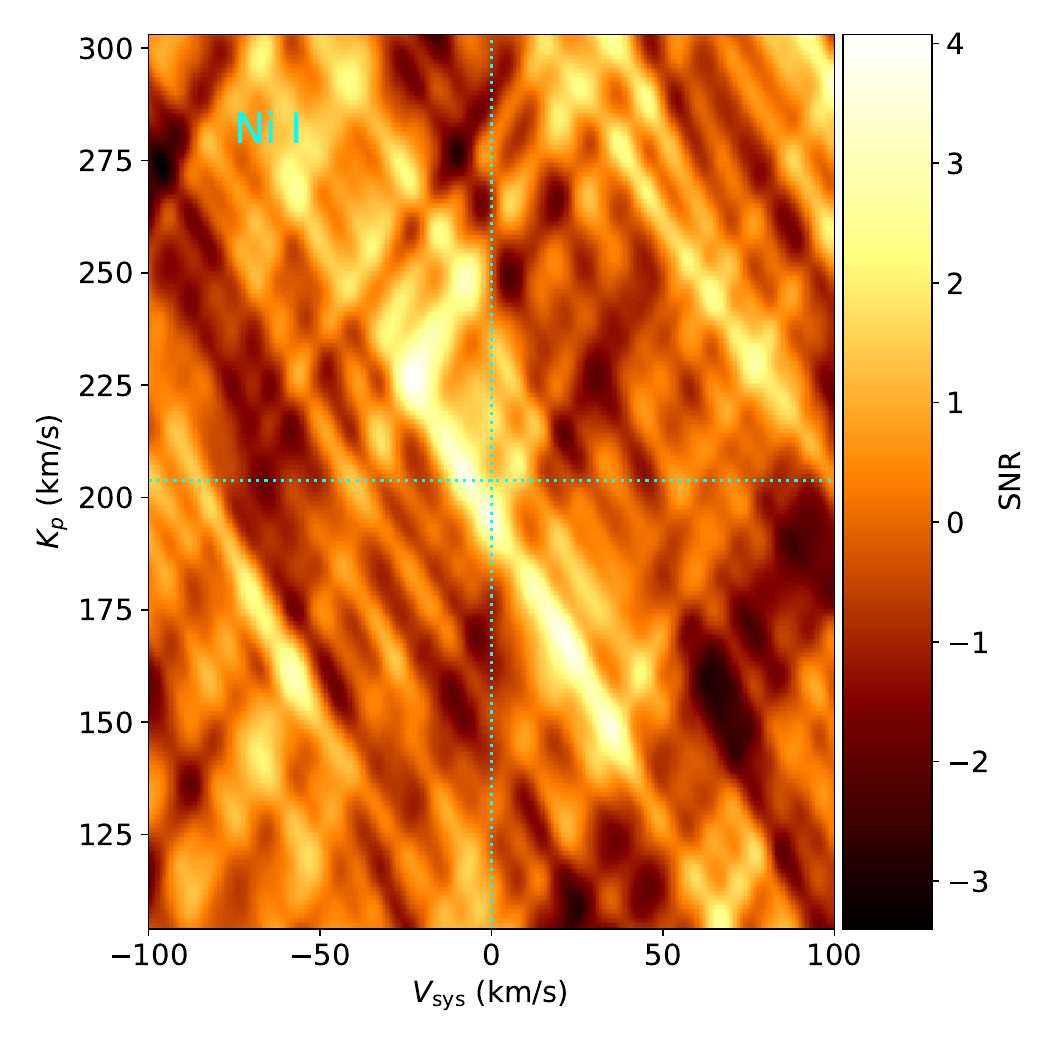}\\
\includegraphics[width=0.4\textwidth]{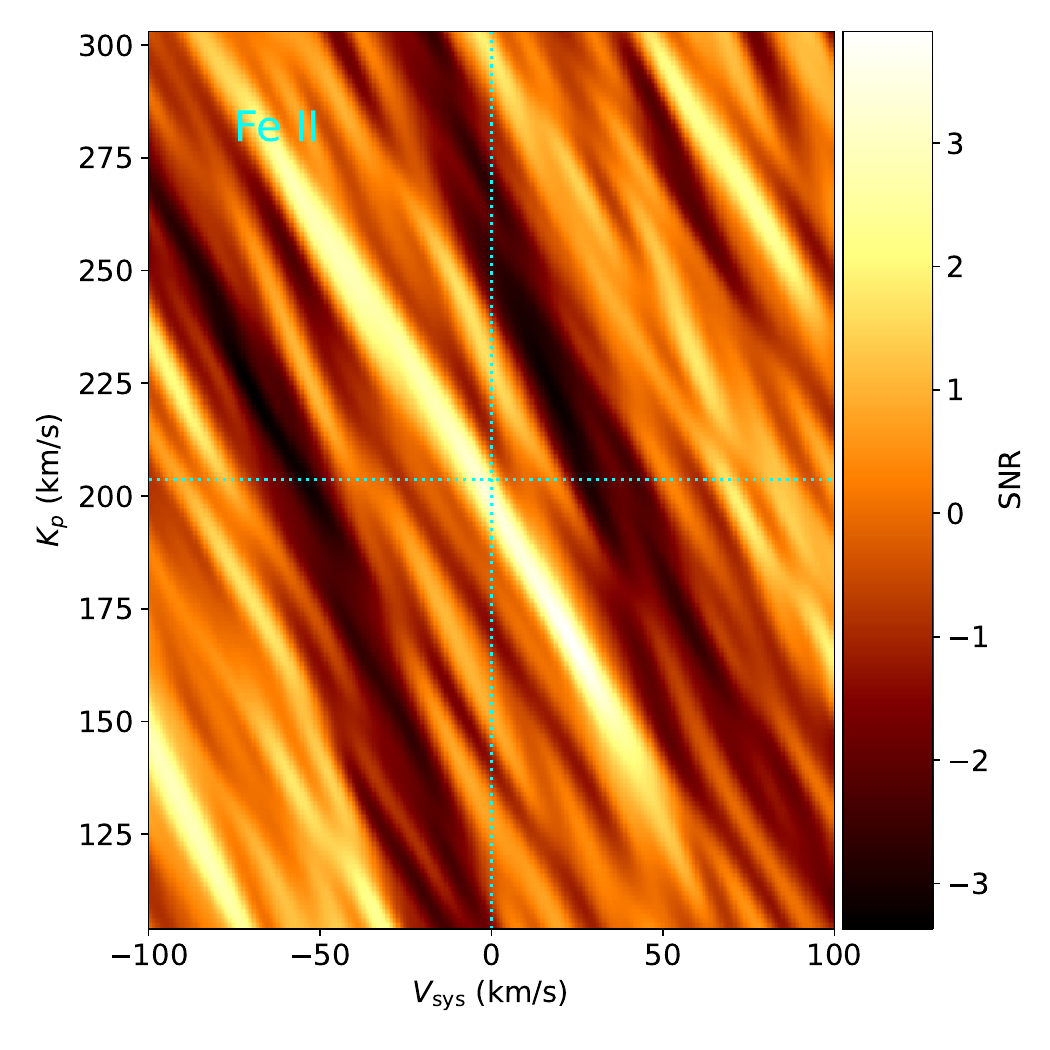}
\includegraphics[width=0.4\textwidth]{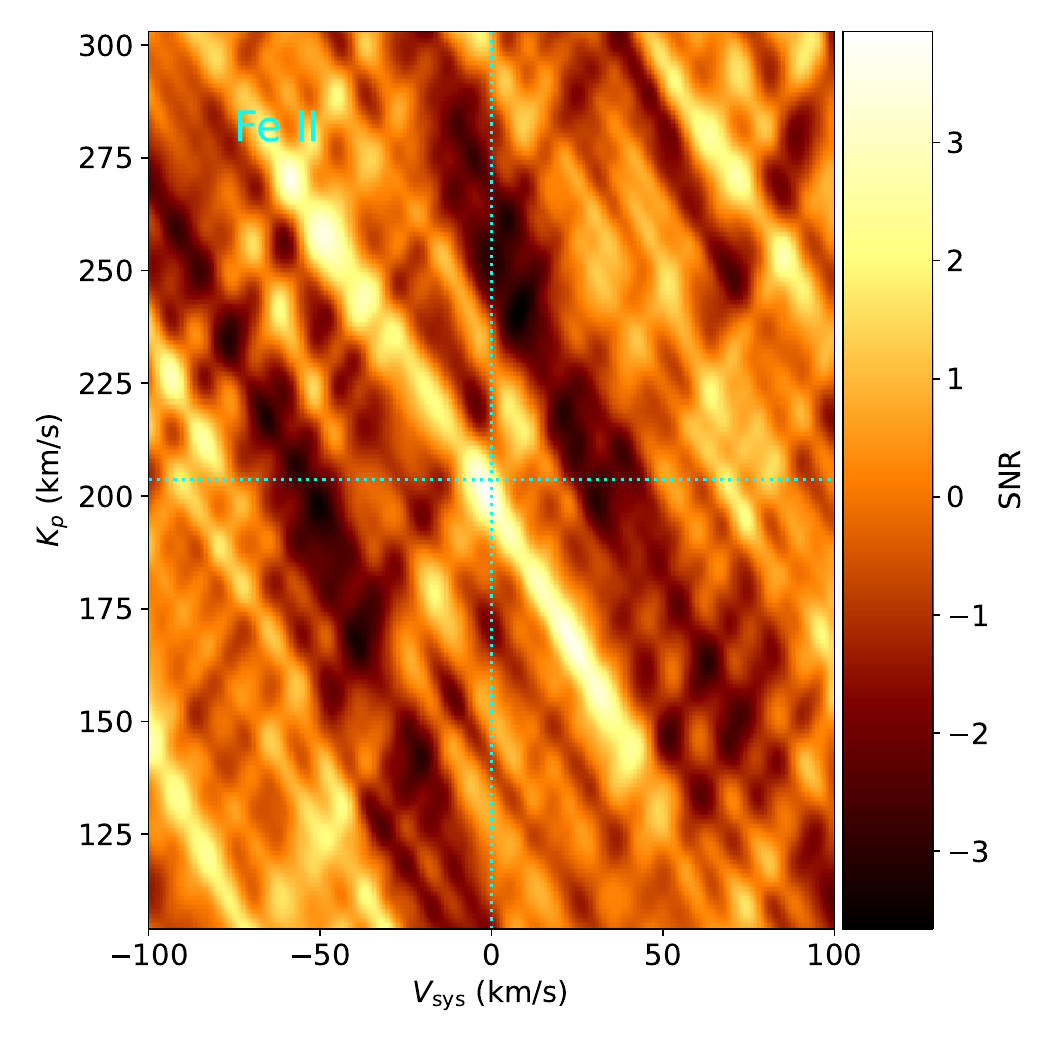} \\
\includegraphics[width=0.4\textwidth]{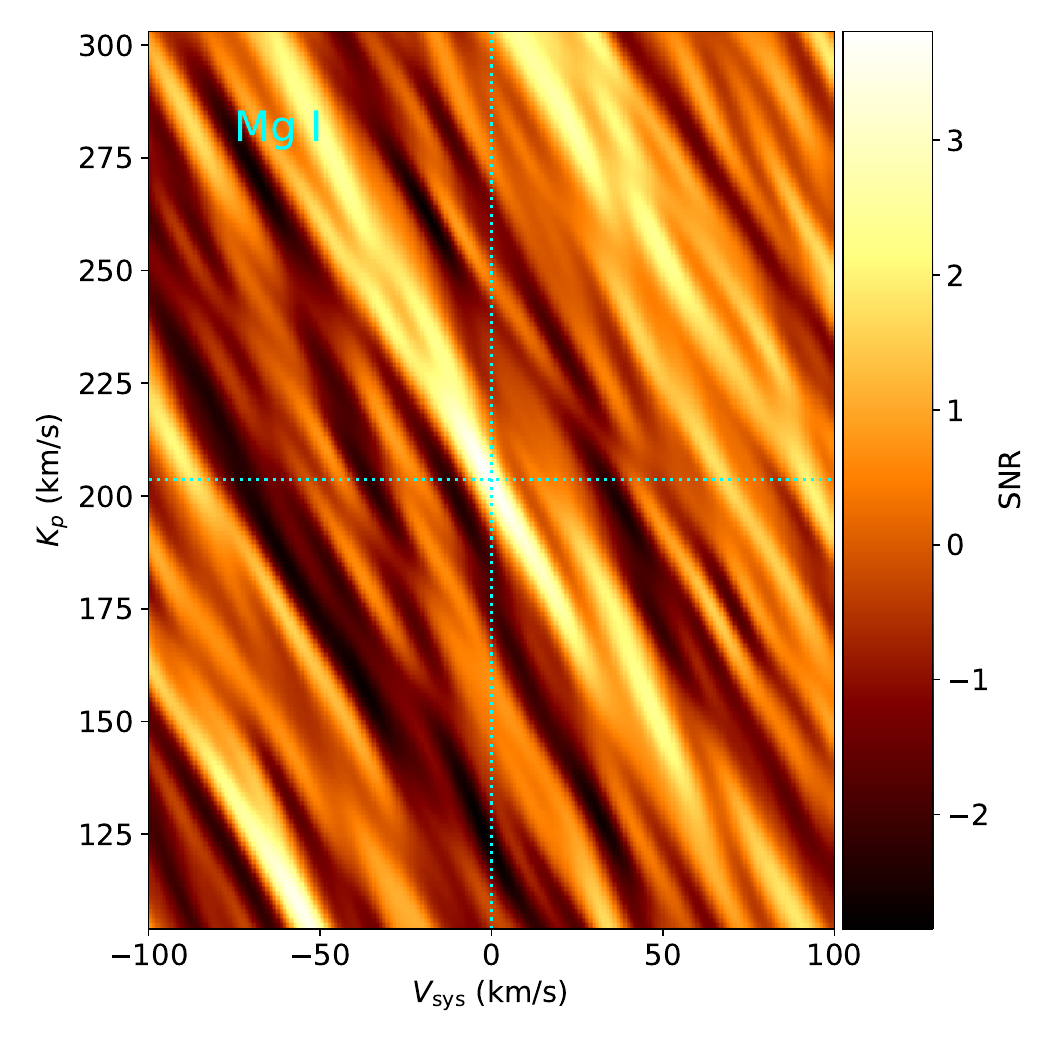}
\includegraphics[width=0.4\textwidth]{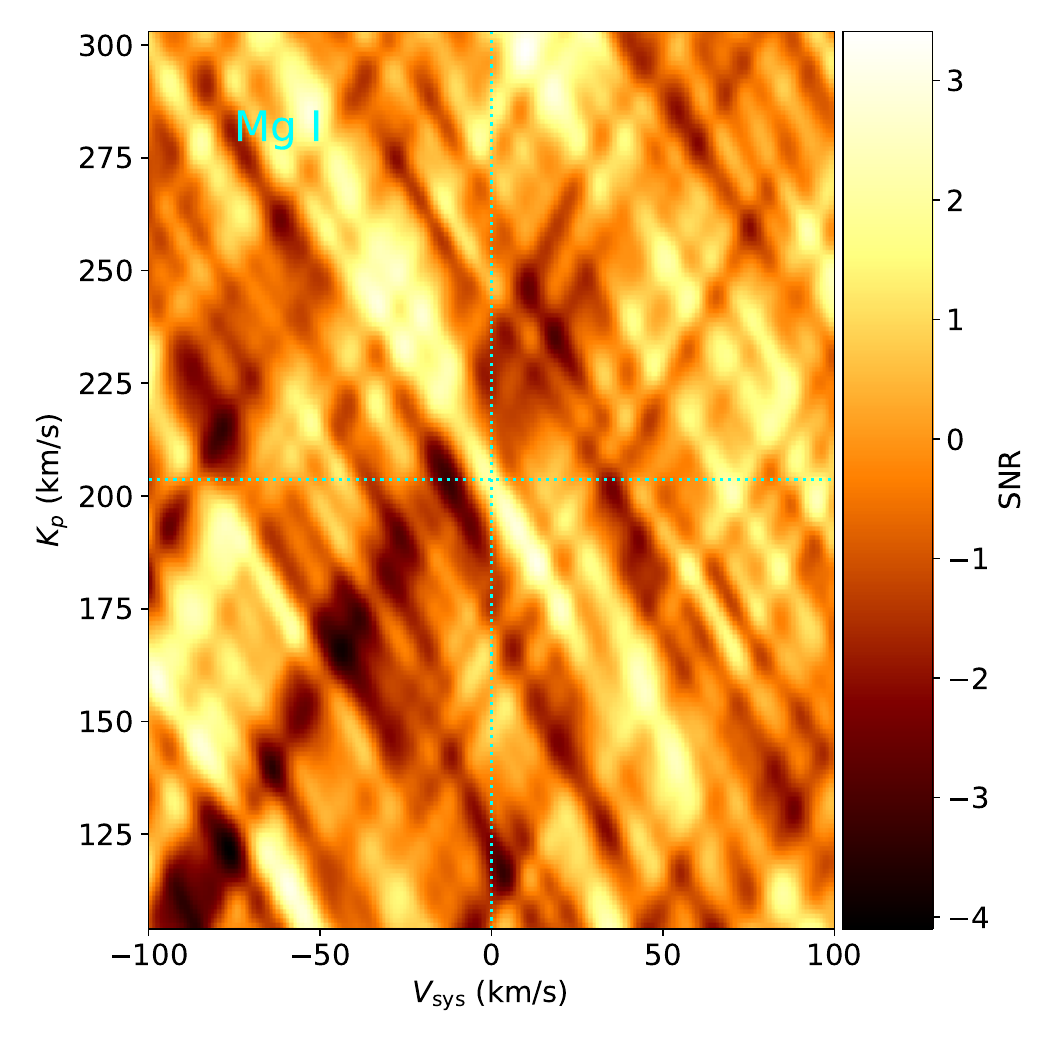}\\
\caption{Shifted and combined CCFs for TOI-1518~b for Ni~\textsc{i} (top row), Fe~\textsc{ii} (middle row), and Mg~\textsc{i} (bottom row) for only the 2023 Nov. 6 observations (left column) and for the combined datasets (right column). The data for Fe~\textsc{ii} use both PEPSI arms, while those for Ni~\textsc{i} and Mg~\textsc{i} use only the red arm. The detection strengths \textit{decrease} in the combined data, perhaps indicating a changing signal strength between the two datasets. See the main text for further discussion.}
\label{fig:ccfs-toi1518-2}
\end{figure*}

We also make tentative detections of Fe~\textsc{ii}, Ni~\textsc{i}, and Mg~\textsc{i} at $3-5\sigma$ significance, shown in Fig.~\ref{fig:ccfs-toi1518-2}. 
The significance of the Ni~\textsc{i} and Fe~\textsc{ii} signals increases only slightly with the addition of the second night of data over only the pre-eclipse observations, while Mg~\textsc{i} drops in the combined data. While the latter trend could be due to 3D effects in the atmosphere changing the signal strength as a function of viewing angle, it could also be due to systematics in the data or other non-physical effects.
We note that for Ni~\textsc{i} and Mg~\textsc{i}, we analyze only the PEPSI red-arm data; the Ni~\textsc{i} lines are concentrated in the red arm, while the Mg~\textsc{i} blue-arm data are dominated by the Mg~\textsc{i} triplet and a strong alias with the Fe~\textsc{i} signal, as previously seen for KELT-20~b by \cite{PETS-IV}. Any Mg~\textsc{i} which might be seen in the blue arm is thus unreliable and we choose to exclude it from the analysis entirely. Combining these three species along with Fe~\textsc{i} opacity into a single model gives an $8.45\sigma$ detection of the atmosphere as a whole in the two nights of data.

Additionally, we see a tentative detection of Cr~\textsc{i} 
at a $3.91\sigma$ significance. There are, however, reasons for caution for this signal. 
The peak of the signal in the 2023 Nov. 6 data (lower left panel of Fig.~\ref{fig:ccfs-toi1518}) occurs at a $K_P$ value approximately $50$ km s$^{-1}$ lower than the expected value. While offsets in the peak $K_P$ between different species have been seen in other planets and ascribed to atmospheric dynamics \citep[e.g.,][]{Brogi23,Cont2024}, the discrepancy for Cr~\textsc{i} is much larger than seen in other planets, raising the possibility that this is not a genuine planetary signal.

We perform a back-of-the-envelope calculation to estimate plausible $\Delta K_P$ amplitudes that could be caused by rotation and atmospheric dynamics, inspired by \cite{Wardenier2023}, who performed a similar calculation for transmission spectroscopy. We will first consider the deviation caused by all emission from the planet coming from the substellar hot-spot, whose RV relative to the planet rest frame changes due to the different projected rotation velocity as the planet rotates. Let $\phi$ be the angle between the line of sight and the line connecting the star and planet, i.e., the phase angle of the planet (assuming a circular orbit). The projected rotational velocity of the hotspot relative to the planet rest frame will then be
\begin{equation}
    \Delta v = v_{\mathrm{eq}}\sin\phi
\end{equation}
Meanwhile, the planetary rest-frame velocity at the same epoch will be
\begin{equation}
    v = K_P\sin\phi
\end{equation}
We now assume that the observations are reasonably close to secondary eclipse, such that we may approximate the sine function as linear. Then, the apparent $K_P$ due to the rotating hotspot is simply
\begin{equation}
    K_{P,\mathrm{eff}} = K_P + v_{\mathrm{eq}}
\end{equation}
from which it is apparent that 
\begin{equation}
    \Delta K_P = v_{\mathrm{eq}}
\end{equation}
that is, the maximal change in $K_P$ due to the planetary rotation is equal to the equatorial rotational velocity (assuming that the planetary rotation is aligned with its orbit). Assuming synchronous rotation, the expected rotational velocity for TOI-1518~b is 5 km s$^{-1}$, insufficient to explain the offset in the apparent Cr~\textsc{i} signal. Similarly, insofar as an equatorial jet manifests as a larger ``rotation speed,'' it would have to have an implausibly large wind speed of tens of km s$^{-1}$ to explain the offset.

\begin{deluxetable*}{lccccccc}
\tablecaption{Summary of Results\label{tab:results}}
\tablewidth{0pt}
\tablehead{
 & \multicolumn{2}{c}{TOI-1431~b} & \multicolumn{4}{c}{TOI-1518~b} & \\
 Species & SNR ($\sigma$) & SNR ($\sigma$) & VMR & SNR ($\sigma$) & SNR ($\sigma$) & VMR & Arms \\
 & both dates & 2023-10-23 UT & & both dates & 2023-11-06 UT &
}
\startdata
All & 8.04 & 8.15 & \ldots & 8.45 & 7.66 & \ldots & combined \\
Fe~\textsc{i} & 5.68 & 5.83 & $5.6\times10^{-5}$ & 7.67 & 6.69 & $3.2\times10^{-4}$ & combined \\
Fe~\textsc{ii} & 2.54 & 2.43 & 1.8$\times10^{-5}$ & 4.05 & 4.00 & $3.2\times10^{-4}$ & combined \\
Ni~\textsc{i} & 2.08 & 1.99 & $3.2\times10^{-6}$ & 4.21 & 4.03 & $3.2\times10^{-5}$ & red \\
Mg~\textsc{i} & 1.36 & 1.63 & $3.2\times10^{-4}$ & 3.39 & 3.62 & $6.8\times10^{-4}$ & red \\
Cr~\textsc{i} & 4.32 & 4.09 & $6.8\times10^{-7}$ & 3.91 & 3.05 & $1.8\times10^{-7}$ & combined \\
\enddata
\tablecomments{The ``both dates'' column lists the peak SNR for the species combining the data from both datasets, while the ``2023 Nov. 6 UT'' and ``2023 Oct. 23 UT'' columns list the SNR only considering data from that date. The quoted VMR is that which maximizes the detection significance and should not be taken as a robust measurement. See the main text for more discussion. ``Arms'' lists which PEPSI arms were included in the combined CCFs. The ``all'' species model combines the species which are reasonably robustly detected for each planet, i.e., Fe~\textsc{i} for both planets plus Cr~\textsc{i} for TOI-1431~b and Fe~\textsc{ii}, Ni~\textsc{i}, and Mg~\textsc{i} for TOI-1518~b.} 
\end{deluxetable*}

\section{Temperature Inversions in the Population of UHJs}
\label{sec:litanalysis}

Temperature inversions appear to be ubiquitous in the dayside atmospheres of UHJs; every UHJ that has been observed to date with high-resolution emission spectroscopy has shown Fe~\textsc{i} and/or CO lines in emission \citep[e.g.,][]{Pino2020,Yan2020,Yan2022,Yan2023,Nugroho2020,Borsa2022,PETS-III,Hoeijmakers2024,Cont2024}. Cooler hot Jupiters, meanwhile, show absorption spectra indicating a monotonically decreasing temperature profile \citep[e.g.,][]{Brogi14,Birkby17,Rasmussen22,Smith2023}. 
With the increasing population of planets observed in emission spectroscopy, we can start to consider population-level questions, including: above which temperature do planets have inverted atmospheres? Is there also a dependence on other parameters, such as surface gravity? What does this tell us about what absorbers are responsible for the presence of inversions?

To begin to answer these questions, we perform a literature search of HJs and investigate their physical properties.
In Table~\ref{tab:litsearch} we list properties from the literature of the sample of HJs with emission spectroscopy indicating the presence or absence of a temperature inversion. This sample of planets consists not only of those observed with high-resolution spectroscopy, but also planets where low-resolution space-based emission spectroscopy with \textit{Hubble} or \textit{JWST} has allowed retrieval of the temperature-pressure profile \citep[e.g.,][]{Mikal-Evans2019,Wong2016}. 

We note that high and low resolution spectroscopy probe different pressure levels in the atmosphere, with higher-resolution spectra probing higher in the atmosphere \citep{Kempton2014, Hood2020, Gandhi2020}. In principle, it might be possible for one to reach different conclusions about the P-T profile from high versus low-resolution data if e.g. there is an inversion high in the atmosphere but a decreasing profile deeper in the atmosphere. For the six planets with both and high resolution spectra, however: WASP-33~b \citep{Haynes2015,Yan2022}, WASP-121~b \citep{Mikal-Evans2019,Hoeijmakers2024}, WASP-18~b \citep{Yan2023,Brogi23}, KELT-20~b \citep{Cont21,Fu22a}, WASP-76~b \citep{West2016,Yan2023}, and WASP-77~Ab \citep{Smith2023,August23}, the two methods agree that the five hotter planets possess temperature inversions, while the cooler WASP-77~Ab does not. One other caveat is that, at high $T_\mathrm{eq}$, a planet with an inverted P-T profile might show a featureless \textit{Hubble} spectrum because this covers the 1.4$\mu$m water feature, which might vanish due to dissociation of H$_2$O and/or continuum opacity due to H$^-$ \citep[e.g.,][]{Parmentier2018}.

Additionally, in order to fill out the low $T_\mathrm{eq}$ end of our sample, we included five non-transiting HJs for which high-resolution infrared spectroscopy has revealed molecular absorption lines and therefore the lack of a temperature inversion: $\tau$ Boo~b \citep{Pelletier21}, HD 179949~b \citep{Brogi14}, 51 Peg~b \citep{Birkby17}, HD 102915~b \citep{Guilluy19}, and HD 143105~b \citep{Finnerty2024}. As these planets are non-transiting, however, we do not know the radius and therefore the surface gravity. Instead, we make use of the relation between irradiation, mass, and radius for HJs to estimate plausible values for their radii. Using the NASA Exoplanet Archive\footnote{\url{https://exoplanetarchive.ipac.caltech.edu/}}, we identify the ten closest transiting planets to each of the aforementioned planets in $T_\mathrm{eq}$-$M_P$ space. Then, we use the mean and standard deviation of the radii of these ten planets as the estimated radius and uncertainty of the planet in question. We list these values in Table~\ref{tab:litsearch}. We also compared these empirically-estimated values to those produced by the \texttt{Forecaster} code \cite{ChenKipping2017}, which uses a single mass-radius relation that marginalizes over irradiation. The radii estimates produced by \texttt{Forecaster} are systematically smaller than ours, but still consistent within $1\sigma$ and thus do not have a significant effect upon our results (which are primarily dependent upon $T_{\mathrm{eq}}$).

\begin{figure*}
\centering
\includegraphics[width=\textwidth]{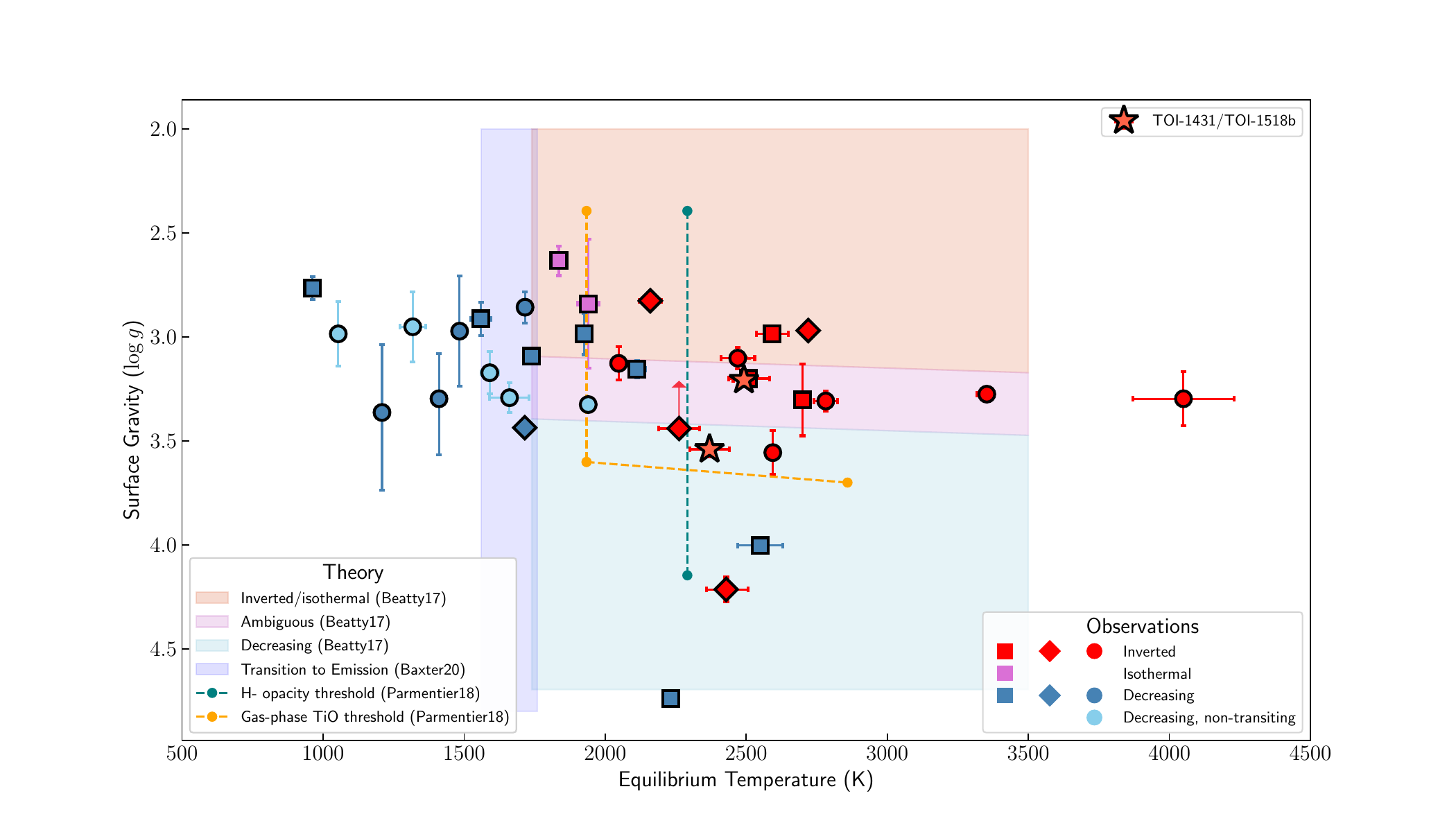}
\caption{Surface gravity as a function of equilibrium temperature of HJs listed in Table~\ref{tab:litsearch}.
Circle, square, and diamond symbols represent planets observed in high resolution, low resolution, and both, respectively.
The orange, purple, and blue shaded regions indicate regions of parameter space proposed by \cite{Beatty17-Kep13} to host planets with inverted or isothermal, ambiguous, and monotonically decreasing P-T profiles, respectively. Note that \cite{Beatty17-Kep13} cast these in terms of the dayside brightness temperature $T_B$ which is a measured quantity not equal to the theoretical quantity $T_\mathrm{eq}$ which we use; however, their relation has only a weak dependence on temperature so the difference should be small. The yellow and teal dashed lines show the boundary found by \cite{Parmentier2018} for the presence of TiO and H$^-$, respectively; these species should be present above and to the right of these lines.}

\label{fig:temp-grav-theory}
\end{figure*}

Following \cite{Beatty17-Kep13} and \cite{Parmentier2018}, we investigate the relationship between equilibrium temperature ($T_\mathrm{eq}$) and surface gravity ($g_P$) with these planets using our sample with a broader parameter space (Figure~\ref{fig:temp-grav-theory}). Surface gravity may be important to the presence of inversions because the chemical agent responsible for the inversions may condense and rain out on the cool night side of the planet, a process which is expected to be more efficient at higher surface gravity \citep{Beatty17-KELT1,Beatty17-Kep13}. \cite{Beatty17-Kep13} proposed a model where the presence or absence of an inversion is dependent primarily on the surface gravity, with only a small dependence upon temperature, at least over the range between 1700 and 3500 K that they considered. \cite{Parmentier2018}, meanwhile, considered the parts of parameter space where gaseous TiO should occur, and where there should be H$^-$ continuum absorption, which has also been suggested to be responsible for inversions \citep{Lothringer18}. These domains are also shown in Fig.~\ref{fig:temp-grav-theory}.

As can be seen in Fig.~\ref{fig:temp-grav-theory}, planets with and without inversion separate into distinct parts of the $T_\mathrm{eq}-g_P$ space. For planets with moderate to low surface gravity $g_P<2 g_J$ ($\log g=3.695$),  every planet with $T_\mathrm{eq}>2150$ K has an inverted atmosphere, while no planet with $T_\mathrm{eq}<1950$ K does. Two planets with isothermal upper atmospheres observed by \textit{Hubble} \citep[HAT-P-31~Ab and HAT-P-41~b;][]{Nikolov2018,Fu2022b} appear at temperatures near 1950 K and low surface gravity. The separation is not completely clean, as with similar surface gravities, KELT-7~b is inverted \citep{Pluriel2020}, while the slightly hotter WASP-19~b is not \citep{Tumborang2024}.
At high surface gravity, the pattern is less clear, and only three objects fill out this space. Kepler-13~Ab and KELT-1~b both have non-inverted atmospheres despite having $T_\mathrm{eq}$ values above 2150 K, while WASP-18~b \textit{does} have an inversion despite having $T_\mathrm{eq}$ and $g_P$ values intermediate between the other two objects.

Comparing our sample to the predictions made by previous literature (Fig.~\ref{fig:temp-grav-theory}), it is clear that with the addition of more recently observed HJs, these predictions do not fully match the current population. A significant number of planets with inverted spectra now occupy the regions predicted by \cite{Beatty17-Kep13} to have ambiguous or absorption spectra, and WASP-18~b shows that the relationship between surface gravity and presence of an inversion is non-monotonic. 
Additionally, WASP-18~b has a surface gravity too high to be expected to have TiO.
We also compare observations to \cite{Baxter2020}, who examined the temperature inversion transition point in temperature space. We find that the transition point in this population at low to moderate surface gravities is somewhat inconsistent with the findings of \cite{Baxter2020} and \cite{Deming2023}, who found transitions in the emission spectra of the HJ population likely due to inversions at $T_{eq} =1660 \pm 100 K$ and $1714-1818$ K, respectively. 
These transitions are at significantly lower $T_{eq}$ than we find; however, they were based upon \textit{Spitzer} secondary eclipse photometry spectrophotometry, which could potentially be due to probing different atmospheric layers at these longer wavelengths, or different viewing geometries for the secondary eclipse data versus the high-resolution phase curves (although the low-resolution \textit{Hubble} and \textit{JWST} data used in Fig.~\ref{fig:temp-grav-theory} would have the same viewing geometry). Further work will be required to resolve this discrepancy.
\cite{Mansfield2021} found a transition from absorption to emission in the \textit{Hubble} 1.4 $\mu$m H$_2$O-band spectra of hot Jupiters at around 2000 K; this is consistent with our results, but with our larger sample we are able to more precisely locate the transition regime and search for trends with gravity in addition to temperature.

To give a more quantitative measurement of where temperature inversions begin to occur, we perform several MCMC fits between the non-inverted and inverted planets in our sample. We use a simple model in $T_{eq}-\log g$ space by constructing a surface such that inverted atmospheres have a value of 1, non-inverted 0, and monotonic 0.5. The regions containing inverted and non-inverted atmospheres are assumed to be separated by a linear function in $T_{eq}-\log g$ space with a sharp transition. We fit this model to the population data using \texttt{emcee} \citep{emcee}. Assuming a temperature only dependent model, we find that the data is best fit by temperature inversions occur at temperatures greater than $2018^{+117}_{-53} K$. Using a temperature and gravity dependent model, we find a stronger dependence on gravity than other previous predictions, with non-inverted atmospheres more likely to occur at higher temperatures if the planet has a higher surface gravity. We use the Bayesian Information Criterion \citep[BIC;][]{Schwarz78} and determine that the $\Delta BIC$ between the two models is $<2$, indicating no significant preference towards either fit. Including versus excluding the three high-gravity objects does not significantly change the results. To further distinguish between both models with any statistical certainty, a larger sample of planets is needed. Additionally, the transition between the two regimes is unlikely to actually be sharp; a larger sample would enable the evaluation of more realistic and sophisticated models, and whether there is a dependence on any additional parameters.

We also compare to more recent models that use the three-dimensional non-grey global circulation model (GCM) grid  presented by \cite{Roth2024}. 
They simulated 345 planets spanning a wide range of parameters that are typical for HJs including temperature and surface gravity, but also including stellar mass, planetary metallicity, magnetic drag strength\footnote{The magnetic drag models are not discussed in \cite{Roth2024} but are included in their data repository at \url{https://zenodo.org/records/10960010}}, and presence or absence of TiO. 
They provided global P-T profiles, spanning 64 longitudinal points, 32 latitudinal points and 53 pressure levels. 
In order to marginalize over the other parameters, not all of which are known for the planets in our observational sample, we simply collapse the grid into $T_\mathrm{eq}-g_P$ space by evaluating the average slope of all the simulated P-T profiles across the entire dayside for each simulated planet between 0.1 mbar and 10 mbar at each $T_\mathrm{eq}-g_P$ grid point.
Following the comparison to observations in \cite{Roth2024}, we choose to implement models that include TiO/VO at $T_\mathrm{eq} >$ 1800K.
 
\begin{figure*}
\centering
\includegraphics[width=\textwidth]{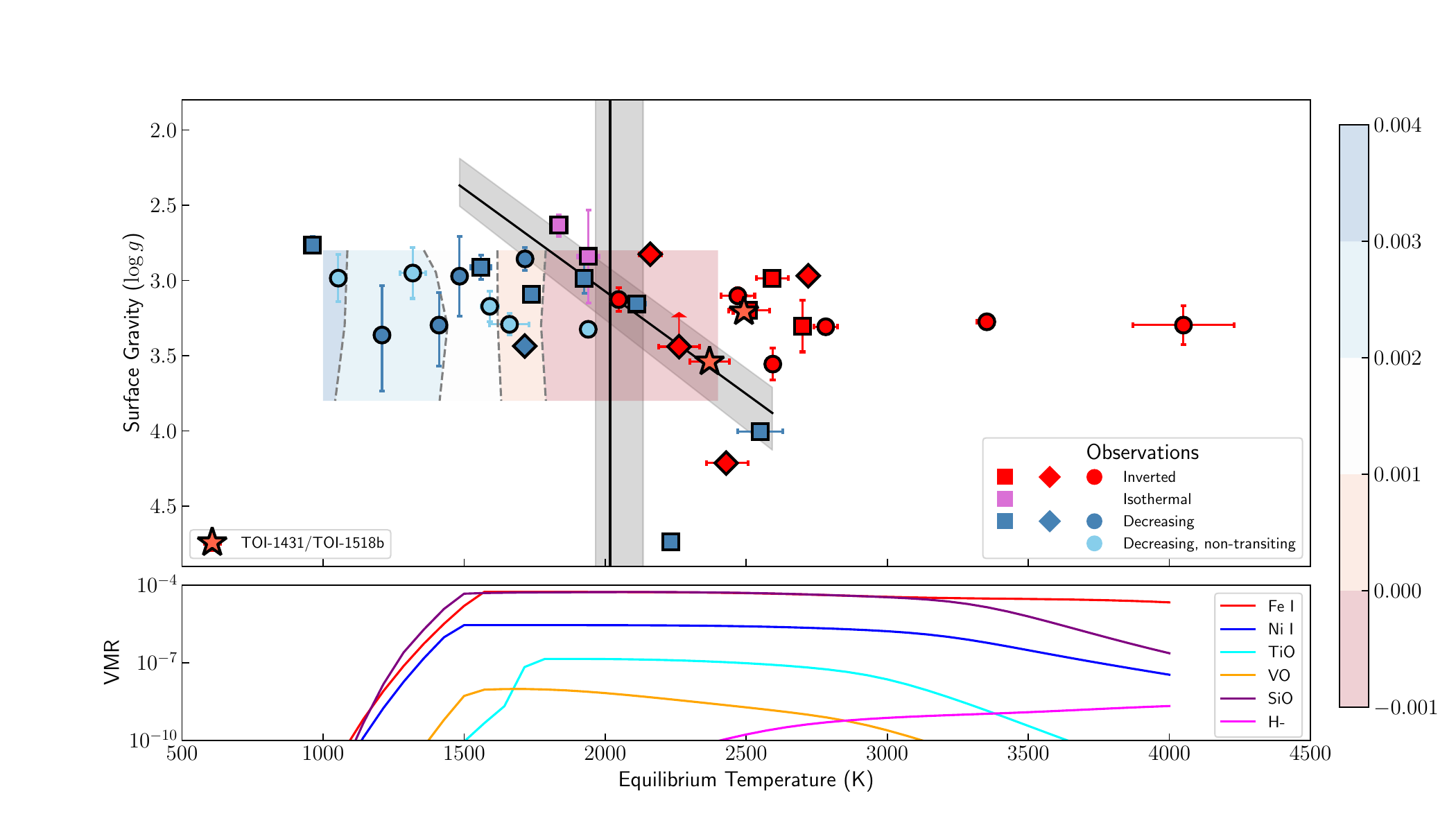}
\caption{Top: same as Fig.~\ref{fig:temp-grav-theory}, except showing contours of P-T profile slope derived from the GCM suite of \cite{Roth2024}. The black dashed line shows the point at which an isothermal P-T profile is expected, while successively more blue-green (red) colors show successively more strongly decreasing (inverted) P-T profiles. The \cite{Roth2024} GCM grid only covers a limited range of $T_{\mathrm{eq}}-g_P$ space as compared to the observed planets. The solid black lines show our best-fit boundary between inverted and non-inverted regimes using a temperature only fit and a temperature and surface gravity fit with a $1\sigma$ shaded uncertainty.
Bottom: expected volume mixing ratios at a pressure level of $10^{-2}$ bar (approximately the pressure level at which the inversions may occur; Fig.~\ref{fig:PTprofiles}) of several proposed inversion agents as a function of planetary equilibrium temperature, as calculated with \texttt{FastChemCond}.} 

\label{fig:temp-grav}
\end{figure*}

In Fig.~\ref{fig:temp-grav} we show contours of this average slope, where negative slopes imply a temperature inversion and positive slopes imply a decreasing P-T profile.
These contours appear to match reasonably well with the observed population with moderate to low surface gravity, predicting the presence of inversions starting at $\sim1800$ K. However, it predicts the transition from non-inverted to inverted profiles at a temperature $\sim250$ K lower than observed. Additionally, the surface gravity range covered by the model grid ($\log g=2.8-3.8$) does not include our high gravity objects like Kepler-13~Ab and WASP-18~b.
The GCM results indicate a minimal dependence on surface gravity. However, this cannot be attributed atomic opacity or condensation processes as the g-dependence of the temperature inversions within the model grid are primarily attributed to variations in heat transfer efficiency.

As an additional test of potential inversion agents, we compute chemical models using the \texttt{FastChemCond} code \citep{Stock18,Stock22,Stock24} in order to compute equilibrium chemical abundances in these atmospheres, accounting for condensation of condensible species. We concentrate on a pressure level of $10^{-2}$ bar, similar to the location where our data suggests the inversions in TOI-1431~b and TOI-1518~b may occur (Fig.~\ref{fig:PTprofiles}). We  compute the expected VMRs of Fe~\textsc{i}, Ni~\textsc{i}, TiO, VO, and H$^{-}$ between temperatures of 500 and 4000 K.

We show the expected VMRs of these species in the bottom panel of Fig.~\ref{fig:temp-grav}. Though it is possible that inversions may be from multiple opacity sources as suggested by \cite{Malik2019} and \cite{Roman2021}, if we believe that a single species is responsible for the inversions, we would expect it to have a significant VMR starting between 1950 and 2050 K, and remaining in gaseous form at higher temperatures. As is apparent from Fig.~\ref{fig:temp-grav}, the best match for this scenario appears to be TiO; Fe~\textsc{i}, Ni~\textsc{i}, and VO would all be expected to be gaseous at the temperatures of known non-inverted planets like WASP-77~Ab and $\tau$ Boo~b, while H$^-$ does not reach significant VMRs until temperatures in excess of 2500 K. 
From this perspective, it seems that TiO may be the most likely cause of the temperature inversions in UHJs. Nonetheless, despite extensive searches over the past decade, observational evidence for the presence of TiO in HJ atmospheres is lacking. \cite{Prinoth22} detected TiO in transmission in WASP-189~b, while potential detections in WASP-33~b have been controversial \citep{Nugroho2017,Cont2021b,Herman2020,Serindag2021}. On the other side, we did not detect TiO in either TOI-1431~b nor TOI-1518~b and set stringent limits on the VMR of $\sim10^{-9}$, two order of magnitude below the expected equilibrium concentration; there are also tight limits for other UHJs \citep[e.g.,][]{PETS-II}. 

Empirically, every UHJ with a temperature inversion observed in the optical shows strong Fe~\textsc{i} emission lines \citep[e.g.,][]{Pino2020,Yan2020,Nugroho2020,Borsa2022,PETS-III,Hoeijmakers2024}. Atomic opacity, including Fe~\textsc{i}, has been proposed as a possible source of inversions \citep{Lothringer18}. While the \texttt{FastChemCond} chemical model for the presence of Fe~\textsc{i} vapor does not match the observed population, we caution that this is a relatively simple, equilibrium chemistry model, and does not account for processes such as atmospheric mixing, photoionization, or the inherently three-dimensional nature of planets \citep[e.g.,][]{Drummond2020,Lee2023,Tsai2024}. 

That Fe~\textsc{i} rather than TiO appears to be the primary source of optical opacity for UHJs with inverted atmospheres suggests that it could plausibly be responsible for the presence of inversions, but we also cannot exclude that species with strong opacity in the UV could play an equal or larger part.
More observations of planets across the parameter space and more detailed atmospheric models are necessary to test the boundaries of the inverted versus non-inverted regimes and to further determine which species are responsible for temperature inversions in UHJs.

\begin{deluxetable*}{lccccccc}
\tablecaption{Temperature Inversions in Hot Jupiters \label{tab:litsearch}}
\tablewidth{0.5\columnwidth} 
\tablehead{
 Planet & HRS & LRS & References & $T_{\mathrm{eq}}$ (K) & $M_P$ ($M_J$) & $R_P$ ($R_J$) & Surface Gravity ($\log g$) \\
}
\startdata
KELT-9~b & Inv & \ldots & 1; 2 & $4050\pm180$ & $2.88\pm0.84$ & $1.891^{+0.061}_{-0.053}$ & $3.297\pm0.130$ \\ 
WASP-189~b & Inv & \ldots & 3; 4 & $3353^{+27}_{-34}$ & $1.99^{+0.16}_{-0.14}$ & $1.619\pm0.021$ & $3.275\pm0.034$	 \\ 
WASP-33~b & Inv & Inv & 4; 5; 57 & $2782\pm41$ & $2.093\pm0.139$ & $1.593\pm0.074$ & $3.308\pm0.048$ \\ 
WASP-121~b & Inv & Inv & 6; 7; 60 & $2720\pm8$ & $1.157\pm0.070$ & $1.753\pm0.036$ & $2.969\pm0.032$ \\ 
HAT-P-7~b & \ldots & Inv & 8; 9; 10 & 2700 & $1.84\pm0.53$ & $1.51\pm0.21$ & $3.303\pm0.172$ \\ 
MASCARA-1~b & Inv & \ldots & 11; 12 & $2594.3^{+1.6}_{-1.5}$ & $3.7\pm0.9$ & $1.597^{+0.018}_{-0.019}$ & $3.556\pm0.105$ \\ 
WASP-12~b & \ldots & Inv & 5; 13 & $2592.6\pm57.2$ & $1.465\pm0.079$ & $1.937\pm0.056$ & $2.985\pm0.035$ \\ 
Kepler-13~Ab & \ldots & Dec & 14; 15 & $2550\pm80$ & $9.28\pm0.16$ & $1.512\pm0.035$ & $4.003\pm0.021$ \\ 
WASP-103~b & \ldots & Inv & 16; 17 & $2508^{+75}_{-70}$ & $1.490\pm0.088$ & $1.528^{+0.073}_{-0.047}$ & $3.200\pm0.034$	 \\ 
TOI-1518~b & Inv & \ldots & 18; This Work & $2492\pm38$ & $<$2.3 & $1.875\pm0.053$ & $<$3.207 \\ 
WASP-178~b & Inv & \ldots & 55; 56 & $2470 \pm 60$ & $1.66 \pm0.12$ & $1.81 \pm0.09$ & $3.102 \pm0.051$ \\ 
WASP-18~b & Inv & Inv & 19; 20; 21; 22; 41 & $2429^{+77}_{-70}$ & $10.20\pm0.35$ & $1.240\pm0.079$ & $4.214\pm0.059$ \\ 
TOI-1431~b & Inv & \ldots & 23; This Work & $2370\pm70$ & $3.12\pm0.18$ & $1.49\pm0.05$ & $3.540\pm0.037$ \\ 
KELT-20~b & Inv & Inv & 24; 25; 42; 43 & $2262\pm73$ & $<$3.382 & $1.741^{+0.069}_{-0.074}$ & $<$3.440 \\ 
KELT-1~b & \ldots & Dec & 26; 27 & 2233.35 & $27.23^{+0.50}_{-0.48}$ & $1.110^{+0.032}_{-0.022}$ & $4.739^{+0.026}_{-0.020}$ \\ 
WASP-76~b & Inv & Inv & 28; 29; 21 & $2160\pm40$ & $0.92\pm0.03$ & $1.83^{+0.06}_{-0.04}$ & $2.826^{+0.032}_{-0.024}$ \\ 
WASP-19~b & \ldots & Dec & 20; 64 & $2113\pm29$ & $1.154^{+0.078}_{-0.80}$ & $1.415^{+0.044}_{-0.048}$ & $3.155^{+0.040}_{-0.042}$ \\
KELT-7~b & \ldots & Inv & 30; 9; 31 & $2048\pm27$ & $1.39\pm0.22$ & $1.60\pm0.06$ & $3.127\pm0.080$ \\ 
HAT-P-41~b & \ldots & Iso & 9; 46; 47 & $1941\pm 38$ & $1.19\pm 0.60$ & $2.05\pm 0.50$ & $2.841\pm 0.310$ \\
WASP-74~b & \ldots & Dec & 9; 53; 54 & $1926\pm 21$ & $0.72\pm 0.12$ & $1.26\pm 0.10$ & $2.985\pm 0.100$ \\
HAT-P-32~Ab & \ldots & Iso & 48; 49 & $1835.7^{+0.110}_{-0.062}$ & $0.68^{+0.11}_{-0.10}$ & $1.98\pm 0.045$ & $2.632^{+0.073}_{-0.068}$ \\
WASP-17~b & \ldots & Dec & 61; 62; 63 & 1738 & $0.512\pm 0.037$ & $1.991\pm 0.91$ & $3.093\pm 0.026$ \\
WASP-79~b & \ldots & Dec & 52 & $1716^{+25.8}_{-24.4}$ & $0.85\pm 0.08$ & $1.53\pm 0.04$ & $2.857\pm 0.075$ \\
WASP-77~Ab & Dec & Dec & 20; 32; 44 & $1715^{+26}_{-25}$ & $1.667^{+0.068}_{-0.064}$ & $1.230^{+0.031}_{-0.029}$ & $3.436\pm0.028$ \\ 
WASP-101~b & \ldots & Dec & 9; 50; 51 & $1560\pm 35$ & $0.51\pm 0.08$ & $1.43\pm 0.09$ & $2.913\pm 0.079$ \\
HD~209458~b & Dec & \ldots & 9; 33 & $1484 \pm 18$ & $0.73 \pm 0.04$ & $1.39 \pm 0.02$ & $2.972\pm0.264$ \\ 
HIP 65~Ab & Dec & \ldots & 45 & $ 1411 \pm 15$ & $3.213 \pm 0.078$ & $2.03_{-0.49}^{+0.61}$ & $3.297_{-0.217}^{+0.271}$ \\
HD~189733~b & Dec & \ldots & 34; 35 & $1209 \pm 11$ & $1.166^{+0.052}_{-0.049}$ & $1.119 \pm 0.038$ & $3.363^{+0.374}_{-0.327}$\\ %Planetary parameters 
WASP-69~b & \ldots & Dec & 9; 58; 59 & $963 \pm 18$ & $0.29 \pm 0.03$ & $1.11 \pm 0.04$ & $2.765 \pm 0.055$\\
\hline
HD~143105~b & Dec & \ldots & 65 & 1940 & $1.23\pm0.10$ & $1.20\pm0.05$ & $3.325\pm0.036$ \\
$\tau$~Boo~b & Dec & \ldots & 36 & $1661 \pm 70$ & $6.24 \pm 0.23 $ & $1.115 \pm 0.092$ & $3.292\pm0.071$\\ 
HD~179949~b & Dec & \ldots & 37; 38 & $1591 \pm 30$ & $0.98\pm0.04$ & $1.28 \pm 0.15$ & $3.172\pm0.101$ \\   
51~Peg~b & Dec & \ldots & 39; 38 & $1318 \pm 46$ & $0.476_{-0.031}^{+0.032}$ & $1.15 \pm 0.22$ & $2.951\pm0.169$ \\
HD 102195~b & Dec & \ldots & 40 & $1054 \pm 13$ & $0.46\pm0.03$ & $1.11 \pm 0.20$ & $2.985\pm0.156$\\ 
\enddata
\tablecomments{Sample of hot Jupiters from the literature used in the analysis described in \S\ref{sec:litanalysis} and shown in Fig.~\ref{fig:temp-grav-theory} and Fig.~\ref{fig:temp-grav}. The sample is listed in descending order of equilibrium temperature. Planets above the line are transiting planets, while those belong the line are non-transiting and we have estimated the planetary radius and surface gravity as described in the main text. The columns ``HRS'' and ``LRS'' denote whether the planet has been observed with high or low resolution, respectively. These observations have resulted in the detection of an inversion (Inv), a decreasing temperature profile (Dec), or an isothermal profile (Iso); an ellipsis denotes no data.
References:  1: \cite{Gaudi2017};  2: \cite{Pino2020};  3: \cite{Lendl2020};  4: \cite{Yan2022};  5: \cite{Chakrabarty2019};  6: \cite{Bourrier2020};  7: \cite{Mikal-Evans2019};  8: \cite{Wong2016};  9: \cite{Stassun2017};  10: \cite{Mansfield2018};  11: \cite{Hooton2022};  12: \cite{Ramkumar2023};  13: \cite{Arcangeli2021};  14: \cite{Esteves2015};  15: \cite{Beatty17-Kep13};  16: \cite{Gillon2014};  17: \cite{Cartier2017};  18: \cite{Cabot2021};  19: \cite{Arcangeli2018};  20: \cite{CortesZuleta2020};  21: \cite{Yan2023};  22: \cite{Coulombe2023};  23: \cite{Addison2021};  24: \cite{Lund2017};  25: \cite{PETS-II};  26: \cite{Siverd2012};  27: \cite{Beatty17-KELT1};  28: \cite{West2016};  29: \cite{Edwards2020};  30: \cite{Bieryla2015};  31: \cite{Pluriel2020};  32: \cite{Smith2023};  33: \cite{Rasmussen22};  34: \cite{Birkby13};  35: \cite{Addison2019};  36: \cite{Pelletier21};  37: \cite{Brogi14};  38: \cite{Rosenthal21};  39: \cite{Birkby17};  40: \cite{Guilluy19}; 41: \cite{Brogi23}; 42: \cite{Cont21}; 43: \cite{Fu22a}; 44: \cite{August23}; 45: \cite{Bazinet2024}, 46: \cite{Hartman2012}; 47: \cite{Fu2022b}; 48: \cite{Nikolov2018}; 49: \cite{Wang2019}; 50: \cite{Hellier2014}; 51: \cite{Rathcke2023}; 52: \cite{Gressier2023}; 53: \cite{Mancini2019}; 54: \cite{Fu2021}; 55: \cite{Hellier2019}; 56: \cite{Cont2024}; 57: \cite{Haynes2015}; 58: \cite{Casasayas2017}; 59: \cite{Schlawin2024}; 60: \cite{Hoeijmakers2024}; 61: \cite{Barstow2017}; 62: \cite{Bonomo2017}; 63: \cite{Gressier2024}; 64: \cite{Tumborang2024}; 65: \cite{Finnerty2024}.}
\end{deluxetable*}

\section{Conclusions}

We have presented high-resolution optical emission spectra of the ultra hot Jupiters TOI-1431~b and TOI-1518~b with LBT/PEPSI (\S\ref{sec:observations}). Using a cross-correlation analysis (\S\ref{sec:methodology}), we detect strong signals of Fe~\textsc{i} emission lines for both planets ($5.80\sigma$ and $7.92\sigma$; \S\ref{sec:bothplanets}), as well as tentatively Cr~\textsc{i} for TOI-1431~b ($4.32\sigma$; \S\ref{sec:1431results}).
We also find tentative detections of $3-4\sigma$ for several other species in TOI-1518~b (\S\ref{sec:1518results}). 

The Fe~\textsc{i} emission lines from TOI-1431~b and TOI-1518~b show that these planets both have temperature inversions in their dayside atmospheres. These add to the ubiquity of both temperature inversions and of optical spectra dominated by Fe~\textsc{i} emission among UHJs at low to moderate surface gravity. We have conducted an analysis of the population of HJs (\S\ref{sec:litanalysis}) by collecting literature values. We display the population as a function of planetary equilibrium temperature $T_\mathrm{eq}$ and surface gravity $g_P$. Previous models proposed by \cite{Beatty17-Kep13} and \cite{Parmentier2018} for the distribution of inversions in this space and transition zones presented by \cite{Baxter2020} in temperature space do not fully match the current population (Fig.~\ref{fig:temp-grav-theory}), but a recent GCM suite from \cite{Roth2024} provides a better match (Fig.~\ref{fig:temp-grav}). While simple equilibrium chemical models from \texttt{FastChemCond} suggest that the distribution of TiO as a function of $T_\mathrm{eq}$ best matches the pattern seen in the population, but this is challenged by the small number of detections of TiO in these planets \citep[e.g.,][]{PETS-II}. Alternately, given that Fe~\textsc{i} emission seems to be ubiquitous in the optical spectra of UHJs observed to date, Fe~\textsc{i} may in fact be responsible for the presence of the inversions, as proposed by \cite{Lothringer18}. However, we cannot rule out the possibility of multiple inversion agents as proposed in \cite{Malik2019} and \cite{Roman2021}. Future emission observations of more HJs in order to probe the presence of inverted atmospheres, or lack thereof, will help to define the limits of the parameter space where inversions occur, as well as further constrain which chemical species may cause the inversions. Particularly needed are planets with intermediate equilibrium temperatures ($1750$ K $<T_\mathrm{eq}<$ 2200 K) and high surface gravity ($g_P>2 g_J$). Even bright non-transiting HJs could be appropriate targets for such observations, particularly with the advent of recent infrared spectra like CRIRES+ and KPIC which are capable of detecting molecules like H$_2$O and CO found across the $T_\mathrm{eq}$ range \citep[e.g.,][]{Cont2024,Finnerty24}.

The PEPSI spectra used in this paper will be made publicly available through the NASA Exoplanet Archive\footnote{\url{https://exoplanetarchive.ipac.caltech.edu/docs/PEPSIMission.html}}.

%% Also note that the akcnowlodgment environment does not support long amounts of text. If you have a lot of people and institutions to acknowledge, do not use this command. Instead, create a new \section{Acknowledgments}.
\begin{acknowledgments}
We thank the LBT support astronomers, telescope operators, and observers who helped to gather our data: Alex Becker, Andrew Cardwell, Steve Allanson, Josh Williams, Mark Whittle, Peter Garnavich, Don Terndrup, and Michael Tucker. Thanks to Scott Gaudi, 
Thomas Beatty, and Vivien Parmentier for useful discussions. Thanks to Jayne Birkby for sharing the improved SYSREM methodology used in this paper. Thanks to the anonymous referee for careful and thorough comments which improved the quality of the paper.

MCJ is supported by NASA Grant 80NSSC23K0692.

The LBT is an international collaboration among institutions in the United States, Italy and Germany. LBT Corporation Members are: The Ohio State University, representing OSU, University of Notre Dame, University of Minnesota and University of Virginia; LBT Beteiligungsgesellschaft, Germany, representing the Max-Planck Society, The Leibniz Institute for Astrophysics Potsdam, and Heidelberg University; The University of Arizona on behalf of the Arizona Board of Regents; and the Istituto Nazionale di Astrofisica, Italy. Observations have benefited from the use of ALTA Center (\url{alta.arcetri.inaf.it}) forecasts performed with the Astro-Meso-Nh model. Initialization data of the ALTA automatic forecast system come from the General Circulation Model (HRES) of the European Centre for Medium Range Weather Forecasts.
\end{acknowledgments}


\vspace{5mm}
\facilities{LBT(PEPSI)}


\software{astropy \citep{2013A&A...558A..33A,2018AJ....156..123A},
\texttt{petitRADTRANS} \citep{Molliere19},
\texttt{FastChemCond} \citep{Stock18,Stock22,Stock24},
\texttt{SpectroscopyMadeEasy} \citep{ValentiPiskunov1996,ValentiPiskunov2012},
SDS4PEPSI \citep{Ilyin2000,Strassmeier18}
          }



%\appendix

%\section{Appendix information}


\bibliography{main}{}
\bibliographystyle{aasjournal}


\end{CJK*}
\end{document}